\documentclass[english,useAMS, usenatbib]{mn2e}
\usepackage[T1]{fontenc}
\usepackage[latin9]{inputenc}
\usepackage{units}
\usepackage{rotating}
\usepackage{url}
\usepackage{amsmath}
\usepackage{commath}
\usepackage{amssymb}
\usepackage{graphicx}
\usepackage[caption=false]{subfig}
\usepackage{pdfpages}
\usepackage{esint}
\usepackage[authoryear]{natbib}
\usepackage{listings}
\usepackage{textcomp}
\usepackage[export]{adjustbox}

\makeatletter

\def\gtsima{$\; \buildrel > \over \sim \;$}
\def\ltsima{$\; \buildrel < \over \sim \;$}
\def\prosima{$\; \buildrel \propto \over \sim \;$}
\def\gsim{\lower.5ex\hbox{\gtsima}}
\def\lsim{\lower.5ex\hbox{\ltsima}}
\def\simgt{\lower.5ex\hbox{\gtsima}}
\def\simlt{\lower.5ex\hbox{\ltsima}}
\def\simpr{\lower.5ex\hbox{\prosima}}
\def\la{\lsim}
\def\ga{\gsim}

\newcommand{\be}{\begin{eqnarray}}
\newcommand{\ee}{\end{eqnarray}}
\def\lsim{\,\lower2truept\hbox{${< \atop\hbox{\raise4truept\hbox{$\sim$}}}$}\,}
\def\gsim{\,\lower2truept\hbox{${> \atop\hbox{\raise4truept\hbox{$\sim$}}}$}\,}
\providecommand{\tabularnewline}{\\}

\title[The history of MW progenitors]{The history of the dark and luminous side of Milky Way-like progenitors}
\author[Graziani et al.]{L. Graziani$^{1}$\thanks{E-mail:
luca.graziani@oa-roma.inaf.it}, M. de Bennassuti$^{2,1}$, R. Schneider$^{2,1}$, D. Kawata$^{3}$, S. Salvadori$^{4}$\\ 
$^{1}$INAF - Osservatorio Astronomico di Roma, Via di Frascati 33, 00078 Monte Porzio Catone, Italy\\
$^{2}$Dipartimento di Fisica, Sapienza, Universit$\grave{a}$ di Roma, Piazzale Aldo Moro 5, 00185, Roma, Italy\\
$^{3}$Mullard Space Science Laboratory, University College London, Holmbury St. Mary, Dorking, Surrey, RH5 6NT,UK \\
$^{4}$GEPI, Observatoire de Paris, PSL Research University, CNRS, Univ Paris Diderot, Sorbonne Paris Cit\'e, \\
      Place Jules Janssen, 92195 Meudon, France}

\begin{document}

\date{November 2016}

\pagerange{\pageref{firstpage}--\pageref{lastpage}} \pubyear{2016}

\maketitle

\label{firstpage}

\begin{abstract}

Here we investigate the evolution of a Milky Way (MW) -like galaxy with the aim of predicting the properties of its progenitors all the way from $z \sim 20$ to $z = 0$. 
We apply \texttt{GAMESH} \citep{2015MNRAS.449.3137G} to a high resolution N-Body simulation following the formation of a MW-type halo and we investigate its properties at $z \sim 0$ and its progenitors in $0 < z < 4$.
Our model predicts the observed galaxy main sequence, the mass-metallicity and the fundamental plane of metallicity relations in $0 < z < 4$. It also reproduces the stellar mass evolution of candidate MW progenitors in $0 \lesssim z \lesssim 2.5$, although the star formation rate and gas fraction of the simulated galaxies follow a shallower redshift dependence.
We find that while the MW star formation and chemical enrichment are dominated by the contribution of galaxies hosted in Lyman $\alpha$-cooling halos, at z > 6 the contribution of star forming mini-halos is comparable to the star formation rate along the MW merger tree. These systems might then provide an important contribution in the early phases of reionization.
A large number of mini-halos with old stellar populations, possibly  Population~III stars, are dragged into the MW or survive in the Local Group. At low redshift dynamical effects, such as halo mergers, tidal stripping and halo disruption redistribute the baryonic properties among halo families. These results are critically discussed in light of future improvements including a more sophisticated treatment of radiative feedback and inhomogeneous metal enrichment.

\end{abstract}


\section{Introduction}

Modern cosmological models \citep{1991ApJ...379...52W, 1994MNRAS.271..781C,2000MNRAS.319..168C, 2015MNRAS.451.2663H} interpret the  properties of galaxies observed in the present 
Universe as the result of the intricate interplay in feedback mechanisms acting during halo 
mass assembly and shaping the galactic baryons through cosmic times \citep{2010gfe..book.....M}. 
Despite the increasing number of objects provided by large scale surveys \citep{1983ApJS...52...89H, 2003AJ....126.2081A, 2015ApJS..219...12A, 2017arXiv170300052B}
and by recent high redshift observations \citep{2011ApJS..197...35G, 2016ApJ...830...67B}, the incomparable 
level of details available at small scale in our Galaxy still places the Milky Way at the 
center of any model of galaxy formation and evolution. 

The possibility of resolving stars, both in the Milky Way and in the closest galaxies 
of the observed Local Group (oLG), provides a unique observational data-set and allows to build Galactic archaeology models 
on solid observational grounds \citep{2006ApJ...653..285S, 2006ApJ...641....1T, 2007ApJ...661...10B, 2007ApJ...658..367K,2010ApJ...708.1398T, 2010MNRAS.401L...5S,2012ApJ...759..115F, 2014MNRAS.445.3039D,
2015MNRAS.447.3892H, 2015MNRAS.454.1320S, 2016arXiv161005777D}.

The great number of kinematic and chemical tracers available in the Milky Way (see 
\citealt{2016ARA&A..54..529B} for a recent review) have been complemented by detailed observations
of stellar populations in nearby dwarf galaxies, enabling us to infer their star-formation histories 
and to interpret the nature of the smallest objects   
(\citealt{2009ARA&A..47..371T,2012AJ....144....4M,2013PASP..125..600M,2016ApJ...819..147M}).
Ultra-faint dwarf galaxies \citep{2007ApJ...670..313S,  2008ApJ...685L..43K, 2014ApJ...796...91B} for example  
are believed to be fossil remnants of the pre-reionization era and the record of radiative feedback at play in the early Universe 
\citep{salvadori09, 2009ApJ...693.1859B, 2009MNRAS.400.1593M, 2014MNRAS.441.2815V, 2015ApJ...807..154B, 2015MNRAS.454.1320S}.
Given this wealth of data, theoretical models have attempted to understand how the star formation history of the
Milky Way and its dwarf companions \citep{2012ARA&A..50..531K} is affected by large scale processes like cosmic 
reionization and metal enrichment \citep{2014ApJ...794...20O, 2015MNRAS.449.3137G,2015ApJ...807...49W}.  

Self-consistent models across scales and cosmic times tuned on low redshift observations have the enormous potential of placing stringent constraints on the nature of unobserved objects at high redshift, as for example the first generation of stars and 
galaxies \citep{2011ARA&A..49..373B}, also helping to investigate the efficiency of early radiative and chemical processes and their mutual impact \citep{2005SSRv..116..625C}.

%%General models of galaxy formation (very short): NBODY: MILLENIUM - MultiDARK - BOLSHOI 
%%and general semi-analytic methods + HYDRO: EAGLE & ILLUSTRIS
Large scale dark matter (DM) simulations, as the Millenium suite\footnote{http://wwwmpa.mpa-garching.mpg.de/millennium/} \citep{2005Natur.435..629S, 2006Natur.440.1137S,2009MNRAS.398.1150B, 2012MNRAS.426.2046A} or the  MultiDark and Bolshoi runs\footnote{https://www.cosmosim.org/} \citep{2011ApJ...740..102K} have traditionally provided the theoretical framework on top of which semi-analytic models interpret global observed quantities (see for instance \citealt{2015MNRAS.451.4029K}).  Recently, two hydrodynamical projects \citep{2015MNRAS.446..521S, 2014Natur.509..177V} including detailed feedback on sub-grid scales have been able to reach an unprecedented realism in reproducing morphological and structural galaxy properties. 

%%Local Group
In the past years many dark matter simulations of MW-like halos and of the oLG have been successfully performed, often as spin-off of large scale simulations, such as the AQUARIUS run \citep{2008MNRAS.391.1685S}, the VIA LACTEA project \citep{2008Natur.454..735D}, the CLUES project \citep{2014NewAR..58....1Y, 2014MNRAS.441.2593N} and the ELVIS simulation suite \citep{2014MNRAS.438.2578G}. They have been extensively used to study the structural properties of dark matter halos, the statistics of their satellites, as well as to correctly constrain the initial conditions leading to the dynamical configuration of our Local Universe, having a MW-M31 galaxy pair \citep{2016MNRAS.458..900C}. Halo assembly histories and the role of mergers events have been investigated as well \citep{2015ApJ...800L...4C}, also finding that no recent major merger should have been occurred during the assembly of our MW \citep{2015A&A...577A...3S}. It should be noted that although very limited in describing the details of baryonic physics, semi-analytic models (SAM) combined with DM simulations are an unavoidable tool to study the statistical properties of galaxy  populations across a broad range of masses and redshifts (see for instance the recent CATERPILLAR project by \citealt{2016arXiv161100759G,2016ApJ...818...10G}).  

Despite a long series of investigations on how to implement mechanical and thermal feedback in different hydro-dynamical schemes \citep{2011MNRAS.414.3671P,2013MNRAS.429..633G, 2016arXiv160707917C} and a  continuous effort in performing code comparison projects \citep{2012MNRAS.423.1726S, 2014ApJS..210...14K}, we are still unable to consistently include feedback processes in models of galaxy formation and evolution. Feedback is often only partially implemented, even not understood in its basic physical principles, and depending on the problem at hand, a theoretical model would favor an accurate treatment of radiation transfer instead of a detailed gas dynamics, being their coupling feasible only under specific conditions. 
In other circumstances, for example when metal ions are used to trace metal enrichment, an alternative, detailed photo-ionization modeling is preferable \citep{2013MNRAS.431..722G}. 
Finally, some observable quantities can be simply more sensitive to photo-ionization than gas dynamics, depending on their physical time scales.

In order to partially compensate these problems, the original version of the semi-analytic, data-constrained model of galaxy formation \texttt{GAMETE} ("GAlaxy MErger Tree and Evolution", \citealt{2007MNRAS.381..647S, 2010MNRAS.401L...5S}) has been considerably extended (see \citealt{2014MNRAS.445.3039D}, hereafter DB14) and self-consistently 
coupled with the radiative transfer code \texttt{CRASH} ("Cosmological RAdiative transfer Scheme for Hydrodynamics", 
\citealt{2001MNRAS.324..381C, 2003MNRAS.345..379M, 2009MNRAS.393..171M, 2013MNRAS.431..722G}) 
creating \texttt{GAMESH}, the first implementation of a full radiative feedback treatment 
in a semi-analitic model on top of a DM only simulation.  
\texttt{GAMESH} is then capable to span the entire formation process of a Milky Way-like halo 
across its cosmic evolution and to target many of the observable properties accessible in the 
local universe (see \citealt{2015MNRAS.449.3137G}, hereafter LG15, for an introduction).

%%GAMETE - GAMESH
This paper is the first of a series where we will progressively exploit the many, new features of the \texttt{GAMESH} model and will  apply it to a number of astrophysical problems, including the reionization history of the Local Universe, the origin and spatial distribution of carbon-enhanced metal-poor stars in the Milky Way halo and its satellites, the formation and coalescence environments of massive black hole binaries, the molecular and dust content of Milky-Way progenitors. Here we present how the \texttt{GAMESH} pipeline has been coupled to a new DM only Galaxy formation simulation performed with the numerical scheme GCD+ \citep{2003MNRAS.340..908K}, obtaining an isolated, Milky Way-type halo in a cosmic cube of about 4 cMpc side length. The application of the semi-analytic model implemented in \texttt{GAMESH} on top of the new N-Body simulation with an increased mass resolution allows us to 
perform an accurate analysis of the Milky Way assembly history through accretion, mergers and dynamical interactions
and to follow the redshift evolution of its baryonic properties, comparing them with observations at $z=0$. 
We investigate the chemical and star formation histories of Milky Way progenitor galaxies and we 
critically compare them with observations of Milky-Way progenitor candidates in the redshift 
range $0.5<z<3$ \citep{2015ApJ...803...26P} and with observed scaling relations in $0 < z < 4$, including
the galaxy main sequence \citep{2015A&A...575A..74S}, the stellar mass-metallicity relation, the fundamental metallicity and fundamental plane of metallicity relations \citep{2010MNRAS.408.2115M, 2012MNRAS.427..906H, 2016MNRAS.463.2002H}.
 
The adoption of a first, simplified feedback scheme allows us to test the reliability of our galaxy formation model across redshift and to discuss the role that different galaxy progenitor populations play in the history of the Milky Way and its Local Group. Furthermore, it allows to easily constrain the main free parameters of \texttt{GAMESH} in order to match the observed properties of the central MW-type halo avoiding complications introduced by the RT effects. Future investigations on the role of radiative and chemical feedback in galaxy formation, especially on low mass galaxies, will adopt the RT described in LG15 and a new particle tagging scheme to simulate inhomogeneous metal enrichment. These features are essential to predict additional observable quantities, such as
the MW metallicity distribution function and the luminosity function of its satellites.

The paper is organized as follows. 
In Section 2 we introduce the new DM simulation and 
we describe the halo catalog, its assembly history and the 
properties of the resulting MW halo. A dedicated Appendix also compares the new N-Body simulation and the properties of the MW halo with four recent, independent simulations: AQUARIUS \citep{2008MNRAS.391.1685S}, ELVIS \citep{2014MNRAS.438.2578G} , CATERPILLAR \citep{2016ApJ...818...10G}, APOSTLE \citep{2016MNRAS.457..844F, 2016MNRAS.457.1931S} (see Table~\ref{table:DMSims}  in Appendix A for more details).

Section 3 describes the tuning of \texttt{GAMESH} on a set of observed Milky Way properties. Section 4 focuses on analyzing the properties of MW progenitors, while Section 5 analyses their evolution as hosted in mini-halos or 
Lyman $\alpha$-cooling halos. Section 6 finally summarizes the conclusions of the paper.    

\section{DM Galaxy formation simulation}

In this section we describe the DM only simulation performed to obtain a MW-size halo. 
We first describe  the numerical scheme adopted in \texttt{GCD+}, the initial conditions of the simulation, the halo catalogue and its merger tree. 
The properties of the MW halo are finally described as well as the statistics of various halo populations in a surrounding volume of 4~cMpc side length.  A careful comparison of the MW properties with similar halos taken from independent simulations both in single and paired configurations can be found in Appendix A, where the halo properties are summarized in  Table~\ref{table:DMHalos} for an easier comparison.

\subsection{ GCD+ and initial conditions}

The N-body cosmological simulation of a MW-sized halo has been performed with GCD+ \citep{2003MNRAS.340..908K, 2013MNRAS.428.1968K} with a $\beta$-version of periodic-boundary conditions and a TreePM algorithm with parallel FFTW module. We used initial conditions created with MUSIC \citep{2011MNRAS.415.2101H} and adopted a Planck 2013 cosmology \citep{2014A&A...571A..16P} ( $\Omega_0=0.32, \Lambda_0=0.78, \Omega_b=0.049$ and $h=0.67$) to simulate a volume of $83.5^3$~Mpc$^3$. 
In this volume we identified a Milky Way-sized halo and we created the initial conditions for a a zoom-in simulation. The final run consists of a total of 62421192 particles, 55012200 of which in
the highest resolution region having particles with mass of $3.4 \times10^5$~M$_{\odot}$. The virial mass of the resulting MW halo is $1.7\times10^{12}$~M$_{\odot}$. 

To better resolve the early evolution of our universe we store the simulation outputs every 15~Myr from $z\sim20$ down to $z=10$, and every 100~Myr after this redshift. The total number of output snapshots is 155. 
This time resolution is high enough to follow the evolution of primordial stellar systems and to correctly account for gas recombinations through cosmic times, when the full pipeline mode including radiative transfer is adopted. 

The final output of the simulation provides a list of collapsed halo objects as well as the projection onto grids of 512$^3$ cells/side of the DM distribution found in the 4~cMpc cosmic volume centered on the MW halo. From these grids the gas distribution in the cosmic web surrounding the MW is easily found by scaling the DM field with the value of the universal baryon fractions indicated by our cosmology. The resulting grid resolution in this domain is then $\sim 7.8$ ckpc. 

\subsection{Halo catalogue}

We identify the populations of DM halos at every snapshot by using a standard friend-of-friends (FoF) algorithm with a linking parameter of $b=0.2$ and a threshold in the number of particles of 100. For each object we stored both virial properties (temperature $T_{\rm vir}$, mass $M_{\rm vir}$ and radius $R_{\rm vir}$), as well the dynamical variables (position and velocity) of its center of mass with respect to the central MW halo. Besides the list of DM halos present in each snapshots, we also stored position and velocities of all particles resolving them. This information allows to study the internal structure  of the most massive halos found in the LG (DM profiles, angular momentum, internal motion, over-density structures, etc..), once they have a sufficient number of particles to reliably compute these quantities. A detailed dynamical study of the MW satellite and sub-satellite properties is still in preparation (see Mancini et al., in prep.).
Before concluding this section, it is worth to note that the initial conditions of the simulation have not been selected to reproduce the structural and dynamical properties of the oLG, but rather to simulate a plausible MW-like halo and to focus on its evolution. As a result, the central 4~cMpc volume contains a total collapsed mass of $M_{\texttt{DM}} \sim 3 \times 10^{12} M_{\odot}$ distributed in $2458$ halos. Among these, 2 halos have DM mass $M \sim 10^{11}$~M$_{\odot}$ (M32, M33 or LMC-type halos, see Table 1 in \citealt{2010MNRAS.404.1111G}  and references therein), 14 have $10^{10} \lesssim M < 10^{11}$~M$_{\odot}$,  and 98 have $10^{9} \lesssim M < 10^{10}$~M$_{\odot}$ (see section~\ref{subsec:MWassembly} for their classification). Also note that the absence of a M31-like halo within 4~cMpc makes the total mass of our MW environment too low with respect to the oLG because $M_{\rm M31} \gtrsim M_{\rm MW}$ by recent estimates (see \citealt{2007ApJ...671.1591I} and references therein.). Other M31-sized halos are found instead in the larger 8~cMpc region. Hereafter we will refer to this 4~cMpc cosmic region as the "LG" of the present simulation. Also note that LG is  also the maximum volume resolved exclusively by the high-resolution DM particles of \texttt{GCD+} and also all its halos are optimally resolved.

\subsection{Merger trees and dynamical interactions}

For each halo found by the FoF at redshift $z_i$, we have built its merger tree (MT) by iteratively  searching all its particle IDs (pIDs) in the previous snapshots, back to the initial redshift $z_1$. A  OpenMP\footnote{www.openmp.org} parallel searching technique, specifically tuned on the simulation data, has been developed to build up the merger tree correlating the pIDs and halo IDs (hIDs) and to establish the ancestor/descendant relationships among hIDs found in $[z_i, z_{i-1}]$\footnote{Note that once a particle belongs to many halos, its multiplicity is also computed and stored to exactly account for the particle contribution in mass transfer processes across the merger tree steps.}. It should be noted that once a pID at $z_{i-1}$ is not associated with any hID, it is associated with a reserved value we call `IGM ID'. We also verified that due to the re-centering adopted for each snapshot to define the LG volume in the simulation data, few pIDs are sporadically not found in the LG volume because not geometrically captured. These pIDs are then classified as `missing' and their associated hID marked as hID$=-1$, to exactly conserve the total mass. It is of primary importance to point out here that in our definition of merger tree the progenitors of a single halo do not necessarily collapse entirely into it, as generally assumed in semi-analytic tree models. In other words, in the merger trees of \texttt{GAMESH} descendant halos do not conserve by design the total mass of their progenitors. The mass is instead conserved when contributions from progenitors are accounted for on a particle base.  

While extremely demanding in term of computational processing, once done, this approach allows us to conserve the mass across dynamical interactions of DM halos with their environments and to exactly follow all their dynamical processes regulating the accretion of dark matter halos: mergers,  tidal stripping and halo disruptions. All these events can then be classified and analyzed and their baryonic counterpart accurately handled in the semi-analytic code. Baryonic properties (generally gas mass, metal mass and stars) can be then properly transferred throughout collapsed structures or returned to the baryonic IGM, by scaling with their relative dark mass contribution. While this is the approach used in the present paper, our scheme is sufficiently accurate to also associate halo baryons to the single DM particles in order to mimic an in-homogeneous spreading in the IGM of the LG by following DM particle dynamics. More details on the particle tagging scheme will be provided in future papers adopting it for specific applications. It should be noted that due to limitations introduced by the mass resolution of our DM simulation and the choice of our FoF parameters to identify halos with 100 particles, our IGM could contain additional population of halos with unresolved masses.        

We classified the various dynamical interactions occurring during the mass assembly in the following categories:

\begin{itemize}  
\item halo growth by accretion: this event occurs when an isolated halo acquires particles only from the IGM, 
typically by mass accretion.
\item Halo growth by merger: when a halo at snapshot $z_{i+1}$ results in a contribution of two or many halos at $z_i$, and possibly the IGM.
\item Halo stripping: when a halo loses part of its mass by tidal interactions with nearby halos.
\item Halo destruction: when a halo found at $z_i$ loses its identity at $z_{i+1}$ because it is disrupted by tidal interactions and its particles are returned primarily in the IGM.
\end{itemize}

In the next section we describe the assembly history of the most massive, Milky Way-like, dark matter halo (hereafter MW) found at the center of the LG cube at $z=0$.

\subsection{The Milky Way DM halo assembly}
\label{subsec:MWassembly}

Here we describe the assembly history of the MW halo defined above, in the context of its Local Group. 

\begin{figure}
\centering
\hskip -1.3truecm
\includegraphics[width=0.55\textwidth]{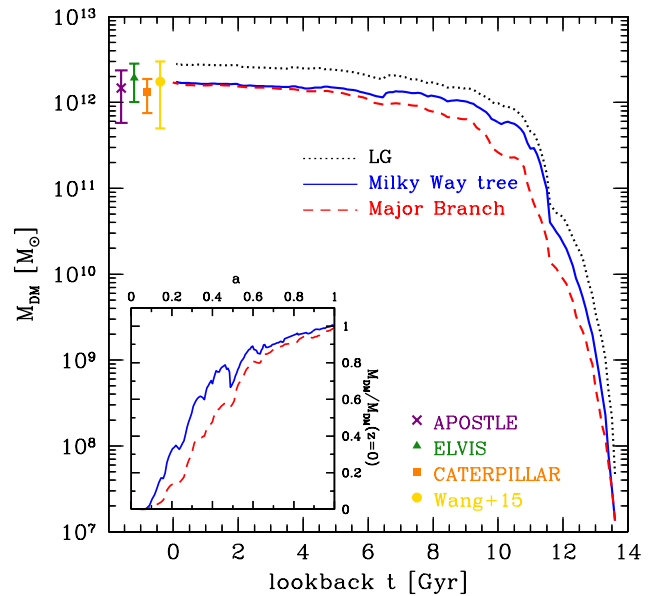}
\vspace{-0.4cm}
\caption{Build up history of the central, MW-sized halo in the adopted N-Body simulation. The total dark matter mass of all the MW progenitors ($M_{\texttt{DM}}$) is shown as function of the lookback time $t$ as solid blue line. The dashed red line shows the MW merger tree as obtained by following the major branch only. The total collapsed mass enclosed in the LG volume is shown as the dotted black line. For reference, the mass of similar MW-sized halos taken from dark matter simulations or independent methods is also shown. Note that the scatter in these values is obtained from the scatter in the MW-halo samples indicated in the various runs (see original papers for details). The inset panel illustrates the same MW history by plotting $M_{\texttt{DM}}(a)/M_{\texttt{MW}}(a=1)$ instead of the total collapsed mass. }
\label{fig:MWHalo}
\end{figure}

\begin{figure*}
\centering
\includegraphics[width=1.0\textwidth]{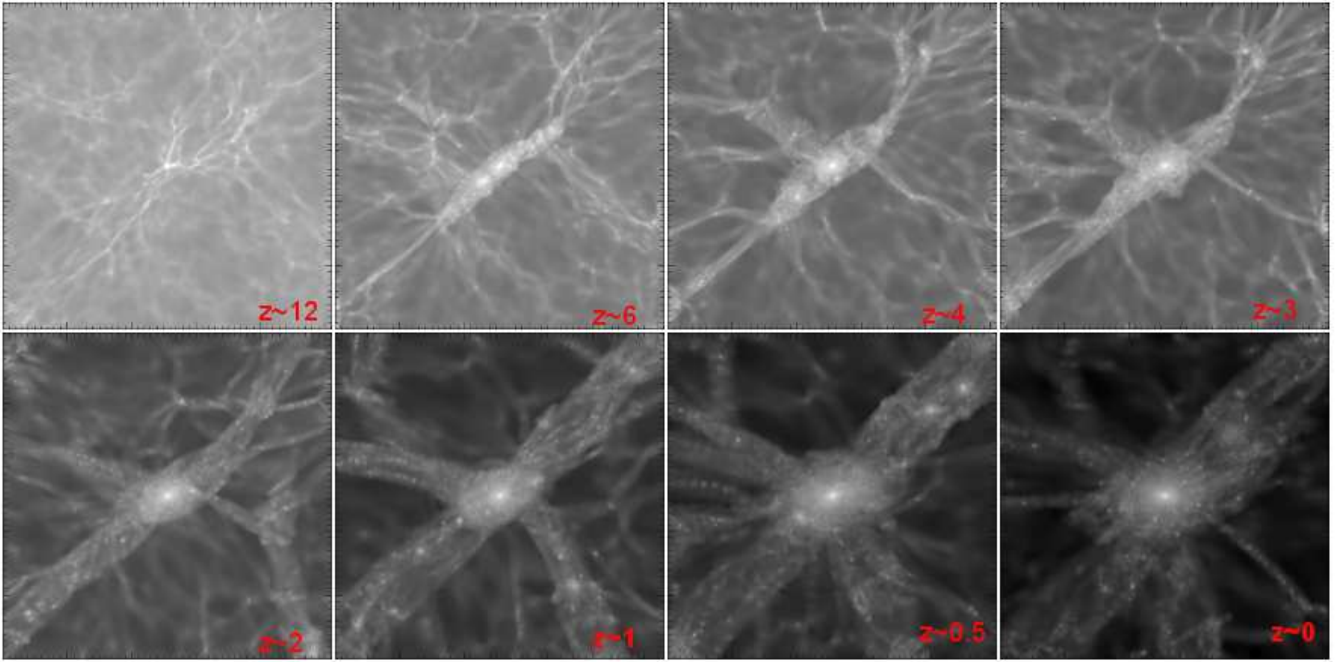}

\caption{Slice cuts of the LG evolution at various redshits. The panels show the DM density map obtaining by projecting the DM mass in each cell of the spatial grid. The total volume is 4~cMpc comoving mapped on a grid of 512 cells/side, for a spatial resolution of $r \sim 7.8$~ckpc.}
         
\label{fig:MWHaloEvol}
\end{figure*}

Figure~\ref{fig:MWHalo} shows the total mass of the MW halo merger tree (solid blue line) as a function of the lookback time ($t$). 
The dotted black line refers to the total collapsed mass found in the LG. 
As described above, our merger trees have been built particle-by-particle and we remind that the blue line shown here accounts, by design, for the total mass of the entire population of progenitor halos providing particles collapsing onto the MW by the successive snapshot. As consequence, the mass shown at certain $z_i$ does not necessarily transfer entirely to $z_{i+1}$ because of the complex series of dynamical interactions at play during the halo mass assembly.  

To highlight the importance of having a complete merger tree we also show, as dashed red line, the MT resulting by following only the mass in most massive halo ($M_{\texttt{MM}}$, or major branch (MB)) at each $z_i$.
It is immediately evident that by following the build up history along the major branch, the discrepancy in mass becomes relevant at high redshift where a sensible fraction of the collapsed mass is distributed in a large number of MW progenitor halos. 
The two MTs converge instead at $t= 4$~Gyr, (i.e. $z\sim0.3$) where a large fraction of the final MW is already collapsed in $M_{\texttt{MM}}$ and all the progenitors contribute a large part of their mass. 
To understand the build up time scale, we computed the so called `characteristic time' ($t_a$) for the assembly of the MW halo, operationally defined as the redshift at which $M_{\texttt{MM}}(z)=M_{\texttt{MW}}(z=0)/2$ (see \citealt{2010gfe..book.....M}) and it results in $t= 4.36$~Gyr, i.e. $z=1.46$. Note that this estimate is compatible with the histories found in an independent set of dark matter simulations described in \citet[see in particular their Figure 19]{2013ApJ...770...57B}, and relative to halos of similar mass. A further comparison with Table~\ref{table:DMHalos} shows that the MW halo has a formation redshift z$_{0.5}$ compatible with the ones of Cat-8 and Cat-12 (both isolated) while the extreme similarity with the formation redshift of Hamilton should be considered as a coincidence because this halo is found in a binary configuration and it is likely to have a very different accretion and dynamical history. Also note that the same time obtained from the blue line (i.e. accounting for all progenitors, see \citealt{1996ApJ...462..563N}) results in $t= 2.66$~Gyr ($z \sim 2.45$), 
further indicating the importance of accounting for the large fraction of mass present as independent collapsed structures at high redshift.

Since in the literature the halo build up histories are usually shown as $M_{\texttt{DM}}(a)/M_{\texttt{MW}}(a=1)$, in the bottom left corner of this figure, we show the same history in these units for a more straightforward comparison with other simulations (see for example Figure 9 in \citealt{2016ApJ...818...10G}).

The resulting MW mass found at $z=0$ ($M_{\texttt{MW}} \sim 1.7 \times 10^{12} M_{\odot}$) is finally compared with the scatter in mass of MW-size halos\footnote{The mass ranges have been computed by extracting min/max mass values in tables of relative papers. Halos in the scatter have been selected as 'reasonably close in mass' to our MW halo and just to suggest an indicative scatter introduced by different methods/simulations. More accurate details can be found both in Appendix A and in the original papers.} found in recent simulations targeting the Local Group, both DM-only (the ELVIS simulation suite by  \citealt{2014MNRAS.438.2578G} and the CATERPILLAR project by \citealt{2016ApJ...818...10G}) and the recent hydro-dynamical APOSTLE simulation \citep{2016MNRAS.457..844F, 2016MNRAS.457.1931S} (see Appendix A for more details). Finally note the additional agreement with the gold filled circle, showing estimates of \citet{2015MNRAS.453..377W}, obtained by using dynamical tracers.  

To understand the MW growth within the global evolution of its LG, one can compare the blue and black solid lines of Figure~\ref{fig:MWHalo} 
and cross-check with the visual picture provided by Figure~\ref{fig:MWHaloEvol}, which shows the redshift evolution 
of the LG in a series of slice cuts intercepting the central MW galaxy. Here the DM density map, is shown 
as gradient from white (collapsed regions) to black (voids)\footnote{Note that the equivalent gas number density is obtained by scaling the DM mass by the universal baryon fraction}.

It is immediately evident that while at high redshifts the mass of both MW and LG have a similar evolution\footnote{This is mainly because 
all the halos collapse first along a filament at the center of the box.}, below $t=11$~Gyr ($z \sim 2$) many structures not belonging to the MW merger tree, start collapsing in the entire volume or enter the domain from the larger scale\footnote{This is a $8$~cMpc cube assumed to  gravitationally constrain the LG domain and contains structures described by lower resolution particles.}. 
The evolution at high redshifts proceeds by assembling halos along the diagonal filament created by the collapsing sheet.
This is easily visible in the first slice cuts (top row, from left to right) where the time evolution of the main web filaments is shown. 
Below $z \sim 3$, the central halo dynamically dominates the LG region and continues to drag material entering from larger scales: around $z \sim 2$, an external filament not previously visible within $4$~cMpc provides halos to the central galaxy. This is a clear hint that galaxy formation  is a multi-scale process, assembling DM/baryonic mass created in different environments along the redshift (see also Section 4). At the final time ($z=0$) the central halo shows a complex interplay with many filaments where a plethora of satellite galaxies are still collapsing towards the central attractor\footnote{An animation can be found in the article online resource files.}. 

\begin{figure}
\centering
\hskip -1.3truecm
\includegraphics[width=0.55\textwidth]{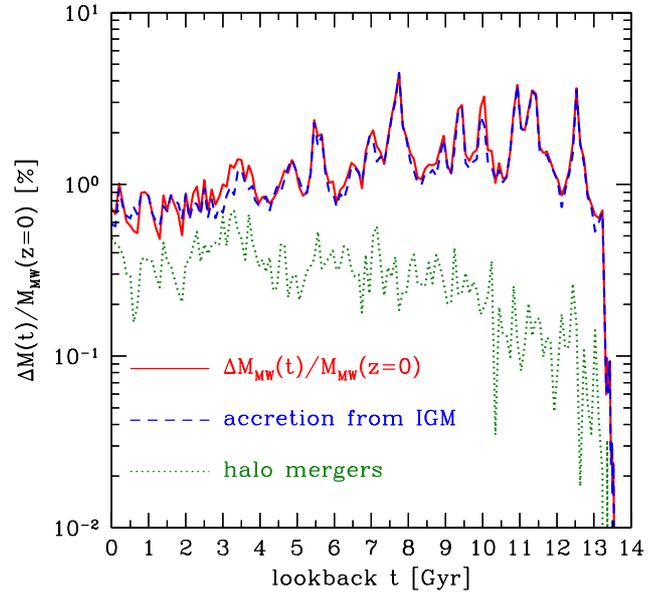}
\vspace{-0.4cm}
\caption{Differential contribution of DM mass, relative to the final MW mass, as a function of the lookback time $t$ (solid red line). The contributions from the IGM and collapsed structures are shown as dashed blue and green dotted lines respectively. }
\label{fig:MWHaloIGMAccretion}
\end{figure}

As explained in Section 2.3, the accuracy of our MT allows us to disentangle the different growth processes (halo mergers or accretion from the IGM) 
and to describe their relative contribution.
The result of this analysis is provided in Figure~\ref{fig:MWHaloIGMAccretion} where we show the percentage of mass increase relative to the final MW mass ($\Delta M(t)/ M_{\texttt{MW}}(z=0)$), as a function of the lookback time $t$ (solid red line). It is immediately evident that across cosmic times, the Milky Way halo grows by means of a 
smooth and continuous assembly of matter spaced out by many violent accretion events, each of which provides a $\sim$3\% contribution to the final mass
(see for example the spikes around $t=12.5$~Gyr ($z=4.68$) and $t=11.5$~Gyr ($z=2.64$)). As a further example note that a major event, 
increasing the mass of the most massive halo by about 5\%, is found around $t \sim 8$~Gyr (more precisely at $z=0.95$), and this also corresponds to the last relevant major merger experienced by two Lyman $\alpha$ (Ly$\alpha$-) cooling halos ($T_{\rm vir} > 2 \times 10^4$~K) found in the MW merger tree\footnote{Note that this is not involving the most massive halo. In fact due to the peculiar history of our MW halo, its last major merger is found instead at $z > 5$.}. Below $t=3$~Gyr ($z \sim 0.2$) both the mergers and the accretion from the IGM phase become smoother and the 
mass growth progresses with steps contributing for less than 1\% to the final mass. Note that minor mergers are continuously found between the MW and small halos or between Lyman $\alpha$ cooling halos and mini-halos ($T_{\rm vir} \lesssim 2 \times 10^4$~K) orbiting the MW.

The relative contribution of accretion and mergers can be understood by comparing the dashed blue line (IGM accretion) and the dotted green 
line (halo mergers). While mass accretion from the surrounding IGM is dominating at all times, the green line shows an increasing number of halo mergers at low redshifts, 
with a substantial contribution around $t \in [2-4]$~Gyr  (i.e. $z \in [0.15-0.35]$). It should be noted though, that while halo mergers contribute on average for less than 
0.5 \% to the final dark matter mass, their stellar, gas and metal contents contribute to shape the observed properties of the MW (see Section 3) and its surrounding satellite galaxies (see Section 4). 

As pointed out in \citet{2011MNRAS.413.1373W}, and also in agreement with the large samples of halo histories found in the Millenium-II simulation \citep{2009MNRAS.398.1150B}, the growth 
of dark matter halos can be dominated by mergers or by a smooth accretion of diffuse matter.  
\citet{2011MNRAS.413.1373W} consider 6 halos of the Aquarius simulation \citep{2008MNRAS.391.1685S} which target a Milky Way-like halo mass at $z=0$ ($1-2\times10^{12}$~M$_{\odot}$) and study their accretion mode and its impact on the internal structure and age distributions of particles in the final Halo.
The authors claim that by averaging over the 6 halos, smooth accretion can provide a relevant contribution to the final mass (roughly 30-40\%). We find  that Aq-A-2 and Aq-C-2  experience a smooth growth history in their major branch, similar to our MW halo. When evaluated with a strict definition of mergers (1:3), the contribution of major mergers to the mass of Aq-A-2 and Aq-C-2 
is below 0.1\%, in agreement with the estimates discussed above for MW\footnote{Note that in this work we adopted a ratio (1:4) to identify a major merger event.}.
Interestingly, their mass and structural properties are also very similar (see details in Appendix A).

In this final paragraph we complement the information provided by the dynamics of the DM matter with the relative contribution of mini-halos and Lyman $\alpha$-cooling halos. In fact, the different impact of radiative and mechanical feedback on these 
populations strongly affects their evolution and leaves imprints on the observed properties of the MW and nearby dwarf galaxies, such as the metallicity distribution function of the most metal-poor stars in the Galactic halo (see \citealt{2010MNRAS.401L...5S}, DB14, LG15, \citealt{2015MNRAS.454.1320S}, and \citealt{2016arXiv161005777D}, hereafter DB16).
   
\begin{figure}
\centering
\hskip -1.3truecm
\includegraphics[width=0.55\textwidth]{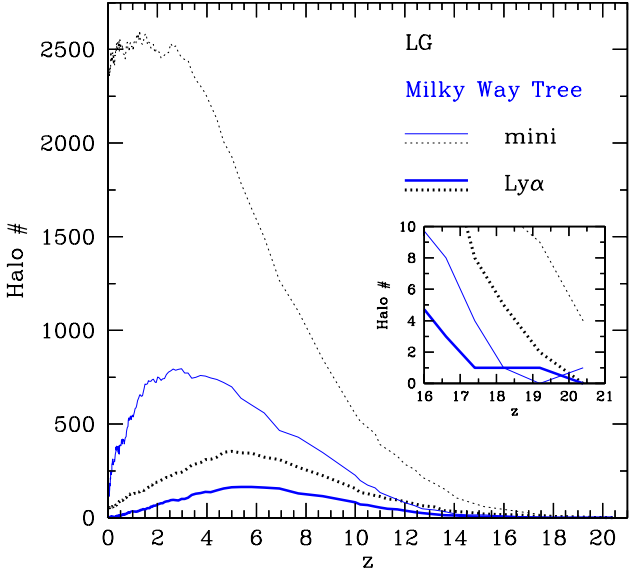}
\vspace{-0.4cm}
\caption{Number of halos found in the LG (black dotted lines) and participating to the merger tree of the MW (blue solid lines) as a function of redshift. 
Mini-halos (Ly$\alpha$-cooling halos) are shown with thin (thick) lines. The enclosed panel shows a zoom-in at high redshift.}

\label{fig:MWHaloPOPs}
\end{figure}

Figure~\ref{fig:MWHaloPOPs} shows the redshift evolution of the number of mini-halos  (thin lines) and Ly$\alpha$-cooling halos (thick lines) 
along the merger tree of the MW (blue solid lines) and in the LG (black dotted lines).  
First note the plethora of mini-halos predicted by the DM simulation around the MW at $z=0$ (N$_{\rm Mini} \sim 550$ within $2\times R_{\rm vir}$). In fact, during the last step of its mass assembly, less than 30\% of the entire mini-halo population in the LG volume is embedded in the final MW halo; these halos remain in the LG, providing a trace of the environmental conditions experienced by this volume along its redshift evolution. While the same considerations apply for the more massive population of Ly$\alpha$-cooling halos, their number is about one order of magnitude lower ($\sim 3$\%), with very few objects (N$_{\rm Ly\alpha} \sim 60$) still orbiting around the MW halo at $z=0$. The role of these two halo populations in shaping the baryonic properties of the central galaxy and its environment will be discussed in Section~\ref{sec:MiniLyaBaryons}.

\begin{flushleft}
\begin{figure*}
\centering
%\hskip -1.1truecm
\includegraphics [trim=1.1cm 0cm 2.2cm 0cm, clip=true, width=5.82cm]{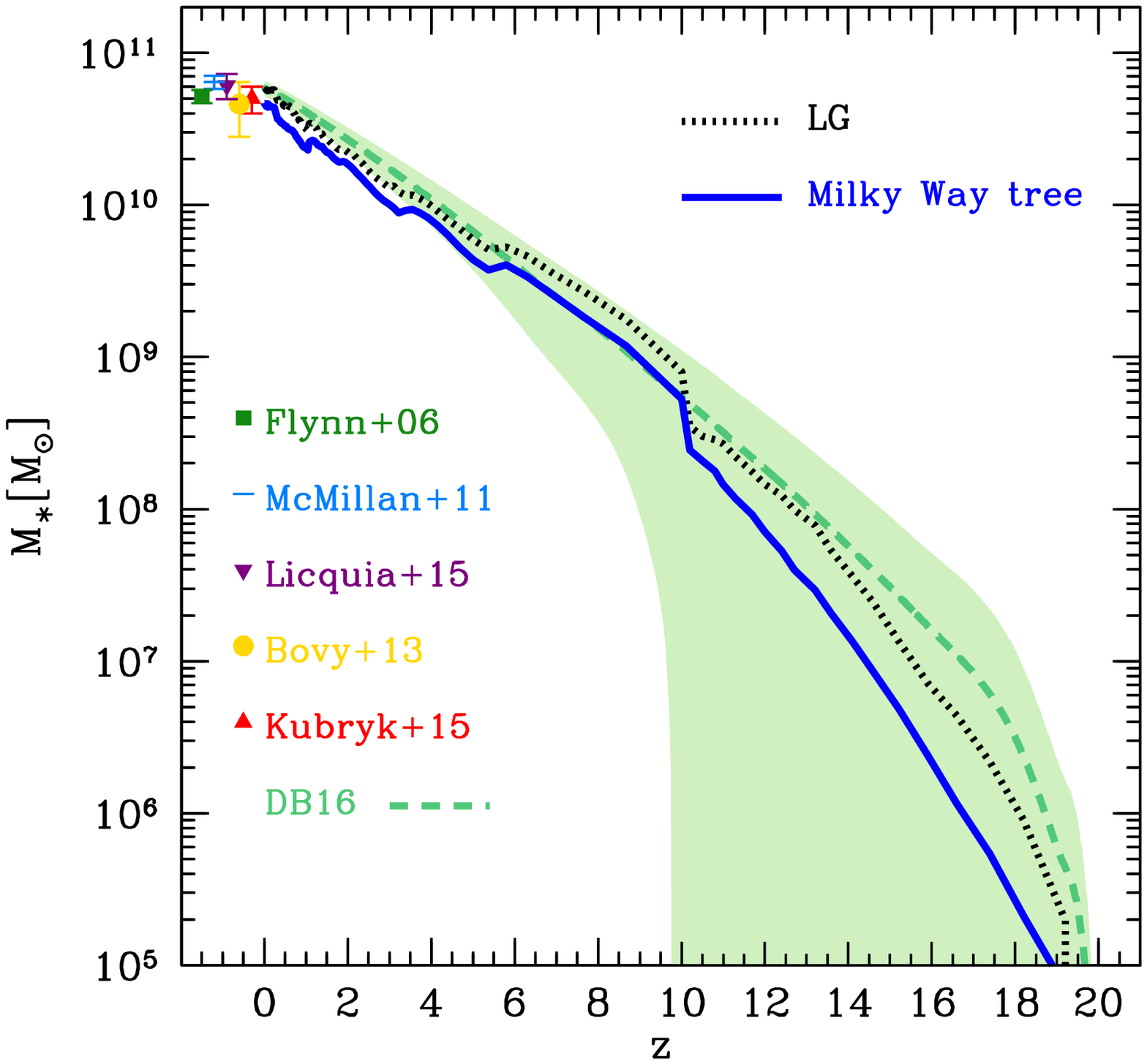}
\includegraphics [trim=1.1cm 0cm 2.2cm 0cm, clip=true, width=5.82cm]{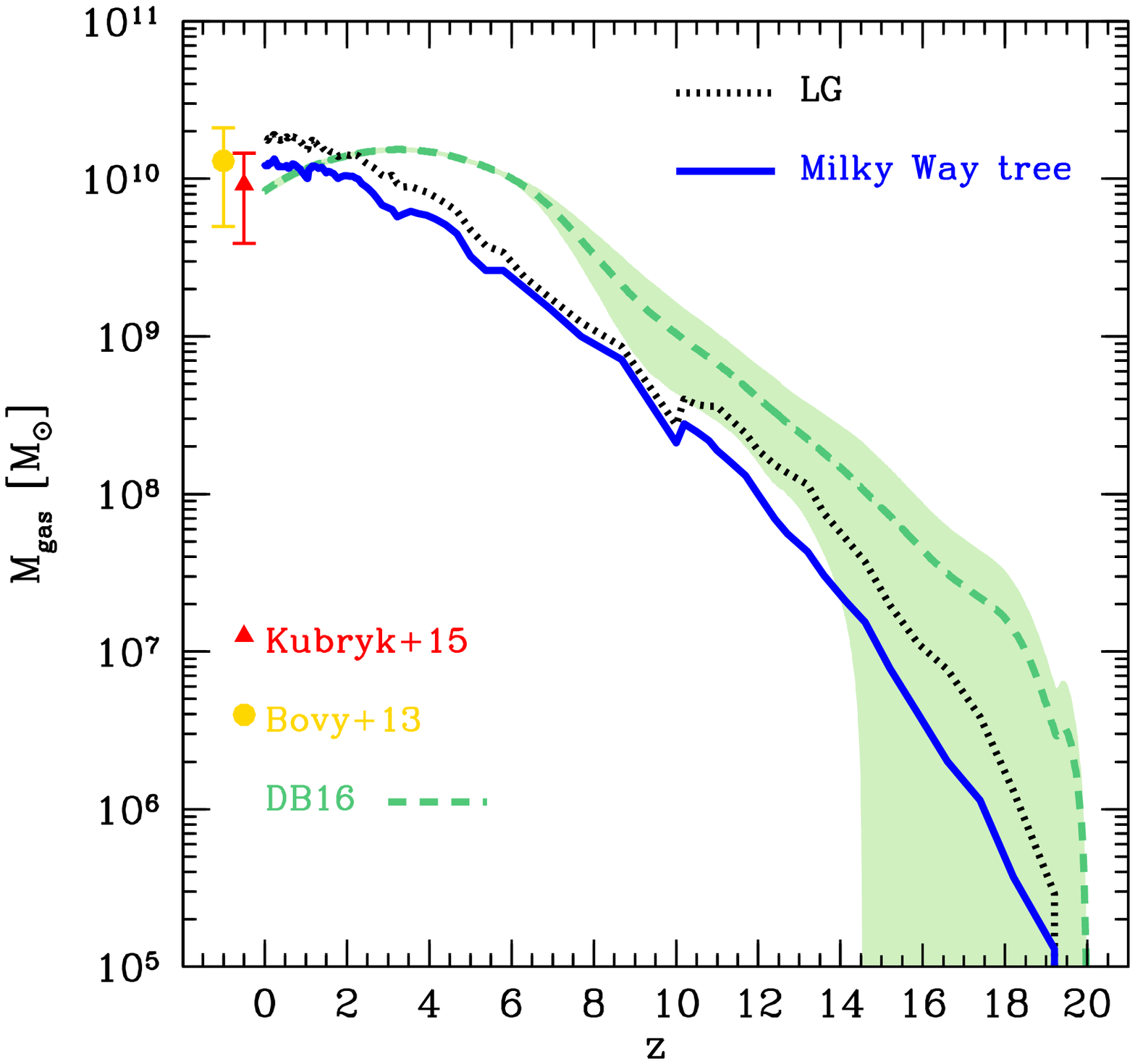}
\includegraphics [trim=1.1cm 0cm 2.2cm 0cm, clip=true, width=5.82cm]{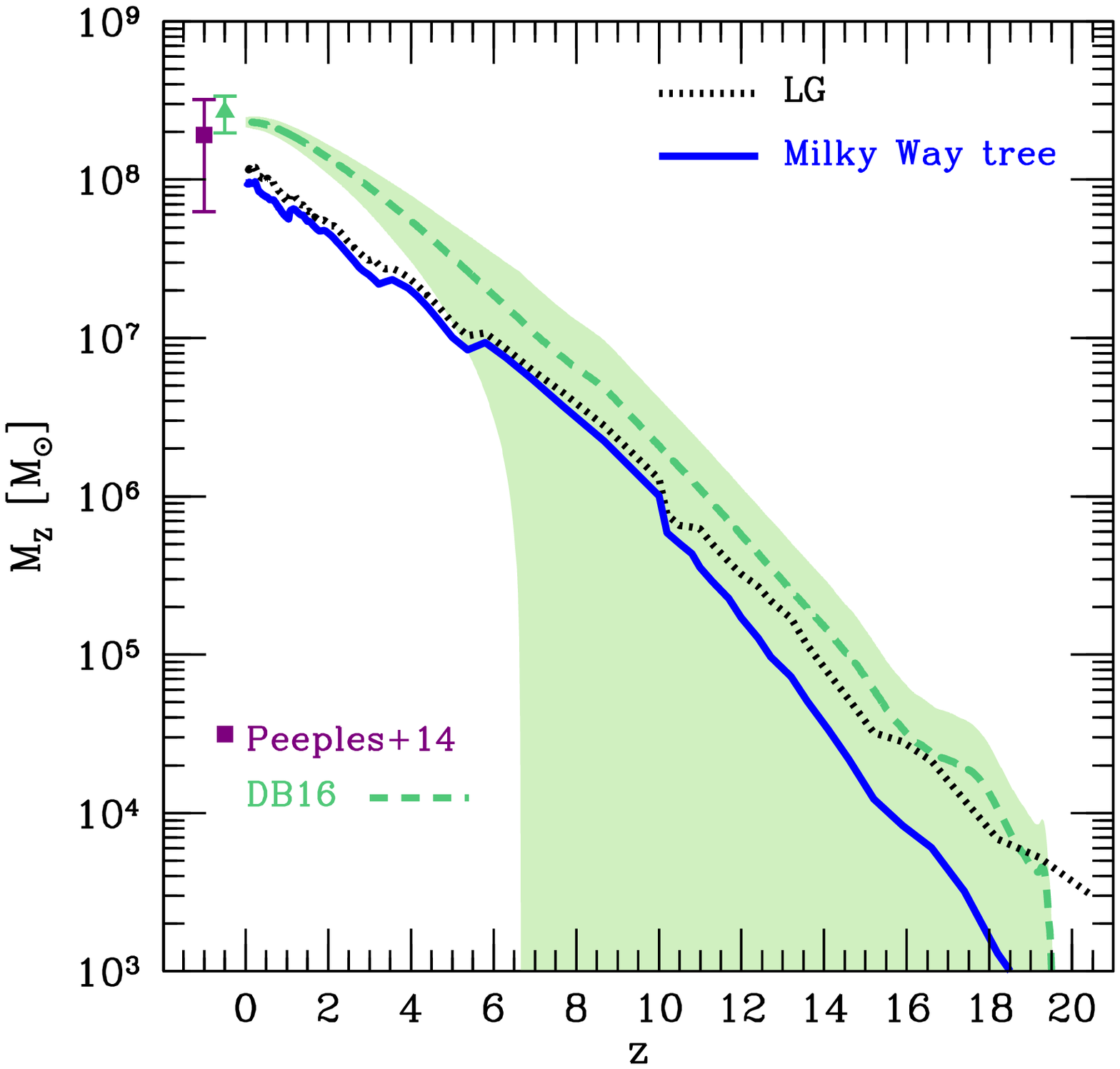}
\vspace{-0.9cm}
\caption{Redshift evolution of the total stellar (left panel), gas (middle) and metal (right) masses. In each
panel, the values predicted by \texttt{GAMESH} for the Local Group and for the merger tree of the MW are shown with black dotted and blue solid lines, respectively.
The dashed green lines show the values computed by DB16 as averages over 50 independent Monte Carlo realizations of a semi-analytical merger tree
for a $10^{12} M_\odot$ dark matter halo (the shaded areas show the corresponding 1-$\sigma$ deviation).
Observations for the stellar and the gas masses at $z=0$ are taken from \citet[green square]{2006MNRAS.372.1149F}, \citet[azure minus]{2011MNRAS.414.2446M}, \citet[violet downtriangle]{2015ApJ...806...96L}, \citet[yellow circle]{2013ApJ...779..115B}, \citet[red uptriangle]{2015A&A...580A.126K}. The derivations of the mass of metals at $z=0$ are taken from \citet[violet square]{2014ApJ...786...54P} and DB14 (light green triangle). Note that in the middle and right panels the blue solid and black dotted lines correspond 
to the mass of gas and metal used in star formation, while the green dashed lines with the shaded regions refer to the total gas and metal mass (see text).}
\label{fig:GAMESH_CALIB} 
\end{figure*}
\end{flushleft}

\section{Baryonic evolution of the MW}

Here we follow the baryonic evolution of the galaxies associated with the DM halos by running the semi-analytic part of the \texttt{GAMESH} pipeline.  
This means that radiative feedback is simulated by adopting a minimum mass of star forming halos and by assuming an instant reionization at $z_{\rm reio}=6$\footnote{Alternative, redshift modulated expressions can be found in \citealt{salvadori09, 2012MNRAS.421L..29S}, DB14, LG15, DB16.}. A separate work will investigate the effects of radiative feedback on star forming galaxies 
in the proper context of a local volume reionization simulation, performed by enabling the \texttt{CRASH} side of the pipeline (see LG15).

While more advanced versions of \texttt{GAMETE} introduce many improvements in the physical processes 
\citep{2012MNRAS.421L..29S, 2014MNRAS.437L..26S, 2014MNRAS.445.3039D,2016arXiv161005777D} or adapt the 
original code to investigate the formation of quasars 
\citep{2011MNRAS.416.1916V, 2014MNRAS.444.2442V, 2016MNRAS.457.3356V, 2016MNRAS.458.3047P},
the pure semi-analytic scheme of \texttt{GAMESH} is based on the following simplifying assumptions:
\begin{itemize}
\item At each given time, stars are formed at a rate given by, ${\rm SFR} = \epsilon_\ast \, M_{\rm gas}/t_{\rm dyn}$, 
where $\epsilon_\ast$ is the star formation efficiency, $M_{\rm gas}$ is the total gas mass, and $t_{\rm dyn}$ is 
the dynamical time of the host halo. 
\item Following \citet{salvadori09, 2012MNRAS.421L..29S}, the star formation efficiency in mini-halos is assumed to be 
$\epsilon_{\rm MH}/\epsilon_\ast = 2 \times [1+(T_{\rm vir}/(2\times10^4 \rm{K}))^{-3}]^{-1}$, as a result of the 
reduced efficiency of gas cooling (see also \citealt{valiante16} and DB16).
\item Stellar evolution is followed assuming the Instantaneous Recycling Approximation (IRA). When the gas metallicity is $Z < Z_{\rm cr}
= 10^{-4} \, Z_\odot$, Population III stars are formed with a constant mass of 200 $M_\odot$. Above the critical metallicity,
Population II stars are formed with masses in the range $[0.1 - 100]~M_\odot$, distributed 
according to a Larson initial mass function (IMF) \citep{1998MNRAS.301..569L} with a characteristic mass of $m_{\rm ch} = 0.35 M_\odot$. 
\item Chemical enrichment by supernovae and intermediate mass stars is based on the same mass- and metallicity-dependent metal yields
adopted in \texttt{GAMETE} and used in LG15.
\item The mass outflow rate of supernova-driven winds is computed as $\dot{M}_{\rm gas, eje} = 2\, \epsilon_{\rm w} \,v_{\rm circ}^{-2}\, 
\dot{E}_{\rm SN}$, where $\epsilon_{\rm w}$ is the wind efficiency, 
$v_{\rm circ}$ the host halo circular velocity, and $\dot{E}_{\rm SN}$ is the energy
rate released by SN explosions, which depends on the star formation rate and on the stellar IMF (hence, a different value
is adopted for Pop~III and Pop~II stars).
\item When $z \leq z_{\rm reio}$ star formation can only occur in galaxies with $T_{\rm vir} > 2 \times 10^4$~K, to account for the effects of photo-heating and 
photo-evaporation (see LG15 for a thorough comparison between the instant reionization model and the model with a self-consistent reionization history computed by 
GAMESH). 
\end{itemize}

We first implemented the simplest version of \texttt{GAMETE} in \texttt{GAMESH} because of many theoretical and practical (mostly numerical) reasons. First, previous runs with \texttt{GAMETE} on a semi-analytic merger tree (see DB16) and on top of a low resolution N-Body (see LG15) have shown that the simplest feedback implementation is sufficient to successfully calibrate the efficiency parameters of the model to fit the main integrated properties of the Milky Way. This in turn,  significantly reduces the range of possible values for our free parameters. Second, the introduction of a higher resolution  simulation affects both the particle scheme of \texttt{GAMESH} and its radiative transfer side, so that the full pipeline becomes numerically demanding even on parallel facilities and then not suitable to make a blind parameters calibration. In future applications showing the new capabilities of the full RT scheme, we will add more observational constrains and will refine the calibration on the new set of observable quantities, also depending on the problem at hand.   

The re-calibration involves the two free parameters of \texttt{GAMESH}, namely the
star formation efficiency in Ly$\alpha$-cooling halos, $\epsilon_\ast$, 
and the efficiency of supernova-driven winds, $\epsilon_{\rm w}$.

As discussed in LG15, we calibrate the free parameters of the model by requiring the star formation rate, the stellar and gas masses,
and the metallicity of the simulated MW galaxy at $z =0$ to match the observationally-inferred values. For some of these quantities,
such as the star formation rate or the total gas mass, the values inferred by different studies show up to one order of magnitude difference, 
as a result of the different tracers used in the observations or of the modeling strategy adopted to reconstruct the galaxy components 
(bulge, disk and halo). The interested reader can find in  \citet{1998ARA&A..36..189K, 2007ARA&A..45..565M, 2012ARA&A..50..531K, 2016ARA&A..54..529B} a large collection of critically revised estimates and galaxy modeling techniques.

Besides the different methodologies, in the last years the total stellar mass ($M_{\star})$ inferred for the Milky Way  has largely converged to a value of 
$M_{\star} = [3-7] \times 10^{10} M_\odot$, as proven by a series of independent estimates \citep{2006MNRAS.372.1149F, 2011MNRAS.414.2446M, 2013ApJ...779..115B,2015A&A...580A.126K,2015ApJ...806...96L}. These are shown by colored points in the left panel of Figure~\ref{fig:GAMESH_CALIB}. 
In the same panel, we show the redshift evolution of $M_{\star}$, as predicted by the more advanced semi-analytic model described in DB16 and ran on top of a semi-analytic merger tree (solid green line with the shaded region) and the stellar mass assembly predicted by  \texttt{GAMESH} in the LG and for the merger tree of the MW, when $\epsilon_{\star}= 0.09$ and $\epsilon_{\rm w}= 0.0016$. The model is in good agreement with the observations, with a final value of $M_{\star} \sim 4.6 \times 10^{10} $~M$_{\odot}$ for the MW halo candidate. 
It predicts a total  stellar mass of $ \sim 6 \times 10^{10} $~M$_{\odot}$ in the  LG and its redshift evolution results consistent with that predicted by DB16 and its statistical scatter. A word of caution is also necessary here when interpreting the evolution in redshift of the baryons accounted for in the \texttt{GAMESH} merger tree and shown in this figure. While \texttt{GAMESH} transfers the baryons from progenitors to descendants exactly scaling by the DM particle contribution, along the merger tree lines shown here the masses do not conserve in redshift, as commented in the DM evolution session. At fixed redshift the mass shown in the merger tree line is the total baryonic mass of the progenitors and not their contribution in mass to the descendants. It is then an estimate of the maximum potential mass available from halo progenitors and the accretion from IGM. Also remember that the mass shown in the LG does not conserve across redshift because of the continuous exchange of systems with the larger scale.   

Below $z \sim 4$ the models show a different evolution, and at $z =0$ the DB16 predicts a mass $M_{\star} \sim 7 \times 10^{10} $~M$_{\odot}$ with 
a local star formation efficiency $\epsilon_{\star}= 0.8$, i.e. one order of magnitude higher than the value required by \texttt{GAMESH}\footnote{Note that in both
models the global star-formation efficiency is defined as, $M_*/M_{{\rm gas}}=\epsilon_* (\Delta t/t_{\rm dyn})$, where $t_{\rm dyn}$ is the redshift dependent halo dynamical time and $\Delta t$ is the time scale assumed for star formation.}. 
The reason for this difference can be ascribed to: ({\it i}) the different mass of the final 
MW dark matter halo, which in DB16 is assumed to be $M_{\rm MW}(z=0) = 10^{12} M_\odot$, 
a factor of 1.7 smaller than the value assumed by the N-body simulation adopted here;
({\it ii}) the different dark matter evolution of the MW halo predicted by the 
N-body simulation, with continuous mass exchanges between halos entering the
MW merger tree and halos of the LG. Conversely, the semi-analytic merger trees
(based on the extended Press-Schechter formalism) are, by construction, mass conserving,
so that any stellar population formed at $z > 0$ along the merger tree will inevitably end up in the MW by
$z = 0$; ({\it iii}) the presence of the IRA that accelerates stellar evolution and underestimates the mass of
active stars present at each given time.

In addition, at $z \leq 4$ the \texttt{GAMESH} MW halo progenitors have
systematically one order of magnitude higher total gas mass than the ones predicted by the semi-analytic merger tree adopted in  DB16. 

This can be seen in the middle panel of Figure~\ref{fig:GAMESH_CALIB}, where we show the redshift evolution of the gas mass.
Here the black dotted and blue solid lines represent the mass of gas used in star formation i.e. $\epsilon_{\ast} M_{\rm gas}$, 
while the green dashed line represents to the 
total $M_{\rm gas}$ predicted by DB16.  As illustrated in Section \ref{subsec:MWassembly} and in Figure \ref{fig:MWHaloEvol},  
the early assembly of the MW halo in the N-body simulation is dominated by mass accretion and mergers of nearby halos, and the evolution
is similar to the one predicted by DB16 using the semi-analytic merger trees based on the Extended Press Schechter formalism (EPS, \citealt{1974ApJ...187..425P}). 
When $z < 4$, the simulated MW-halo grows by many episodes of violent accretion and many minor mergers with halos entering the LG from the larger scales. All these effects cannot be accounted for by the semi-analytic merger trees.

Despite the intrinsic differences found in their assembly histories, both models predict a final gas mass in the MW in agreement with the observed values.
Here the comparison among models and with observations should be taken with caution. In fact, the total gas mass in the MW includes the cold and warm components 
(molecular and atomic phases mainly in the disk) and a hot halo (coronal) component, as exhaustively detailed in \citet{2016ARA&A..54..529B}.
The observations reported in the middle panel of Figure~\ref{fig:GAMESH_CALIB} refer to the total interstellar medium (ISM) mass in the disk as
inferred from dynamical measurements \citep[yellow circle]{2013ApJ...779..115B}, and by averaging the values of the atomic and molecular gas masses 
obtained by different observational studies \citep[red triangle]{2015A&A...580A.126K}. We note that by adopting the most likely mass range for the Galactic corona 
$(2.5 \pm 1) \times 10^{10} M_\odot$, \citep{2016ARA&A..54..529B}, the total baryonic mass is estimated to be in the range 
$[7 - 11]\times 10^{10} M_\odot$. If we account for the total amount of gas enclosed in the MW halo, we find 
$M_{\rm gas} \sim 1.3 \times 10^{11}M_\odot$. 

\begin{figure}
\centering
\hskip -1.3truecm
\includegraphics[width=0.50\textwidth]{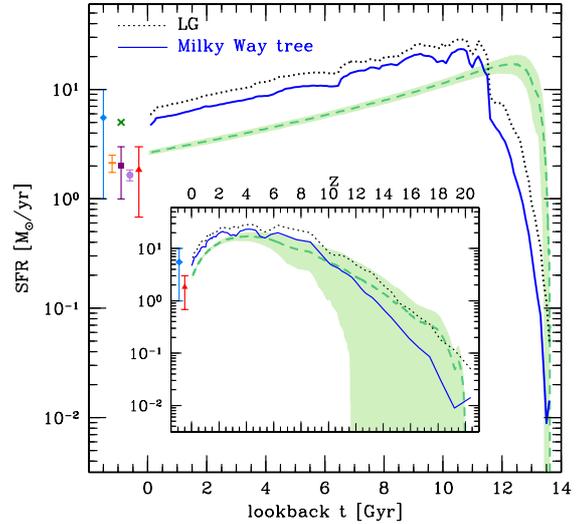}
\vspace{-0.4cm}
\caption{Total star formation rate in the Milky Way merger tree (solid blue line) and in the LG (dotted black line) as function of the lookback time. The average SFR found in DB16 is shown with dashed green line, with the
shaded region showing the 1$-\sigma$ dispersion. In the enclosed panel the same quantities are shown as function of $z$. Data points with errorbars, when available,
 are taken from the literature (see text for details).}
\label{fig:GAMESH_CALIB_SFR}
\end{figure}

Finally, in the right panel of Figure~\ref{fig:GAMESH_CALIB} we show the evolution of the metal mass in the ISM. Here we do not follow separately
the evolution of dust, hence all the lines show the total mass in heavy elements (gas-phase metals and dust). The models show 
behaviors which reflect their corresponding stellar mass assembly histories, and DB16 predicts a higher metal content
at $z=0$, consistent with its higher $M_{\star}(z=0)$. Both models are in agreement with the violet square at $z=0$,
based on a detailed inventory of metal mass components in present-day $L_\ast$ galaxies \citep{2014ApJ...786...54P}.
To be consistent with the observations reported in the other panels of Figure~\ref{fig:GAMESH_CALIB}, we have computed the mass of metals and dust 
in the ISM from the fitting functions of \citet{2014ApJ...786...54P}, using a stellar mass in the range $[3-7]\times 10^{10} M_\odot$ and we find
$M_{\rm Z} = [0.95-4.7] \times 10^8 M_\odot$.

\begin{figure*}
\centering
%\hskip -1.3truecm
\includegraphics[width=0.32\textwidth]{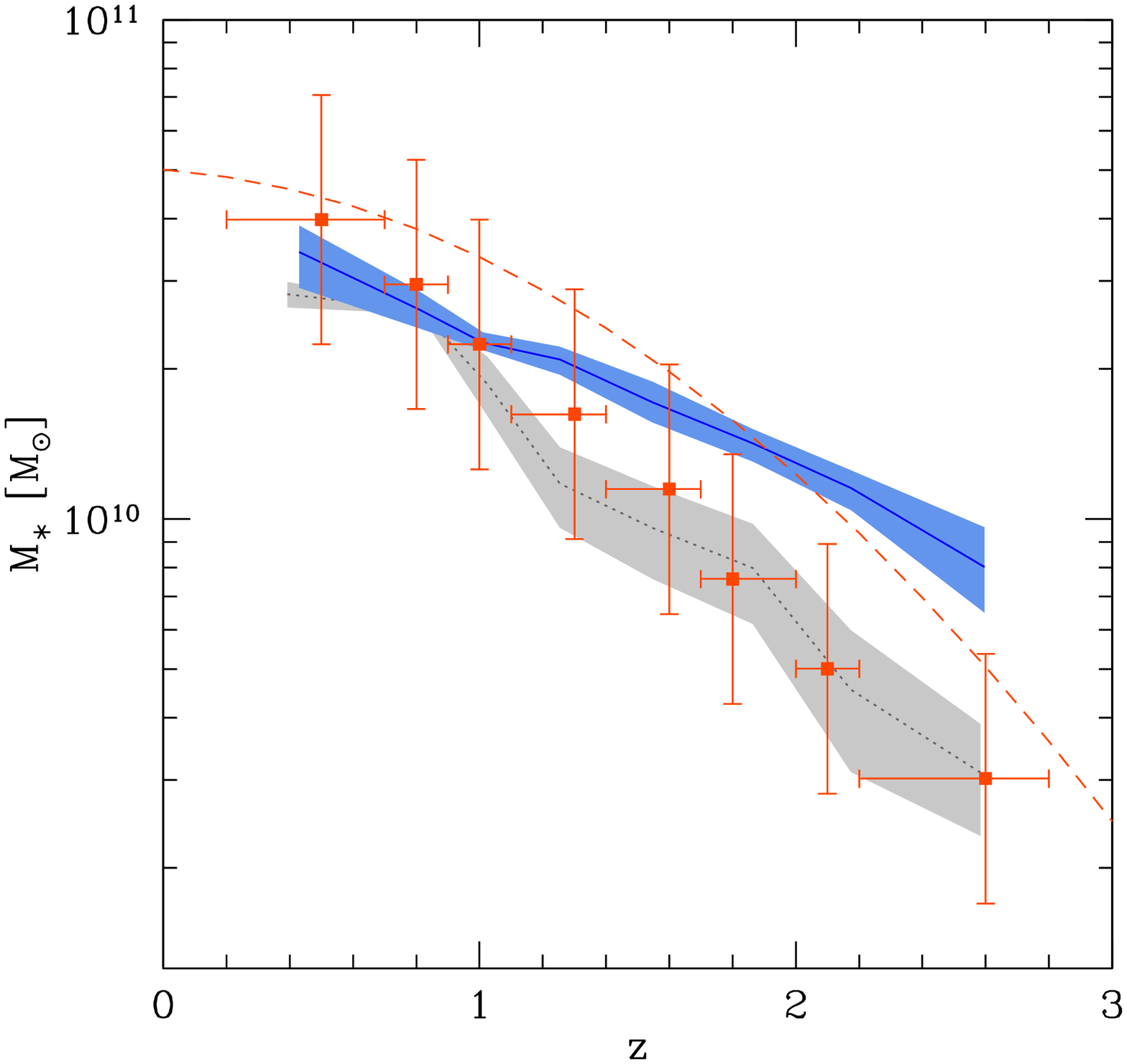}
\includegraphics[width=0.32\textwidth]{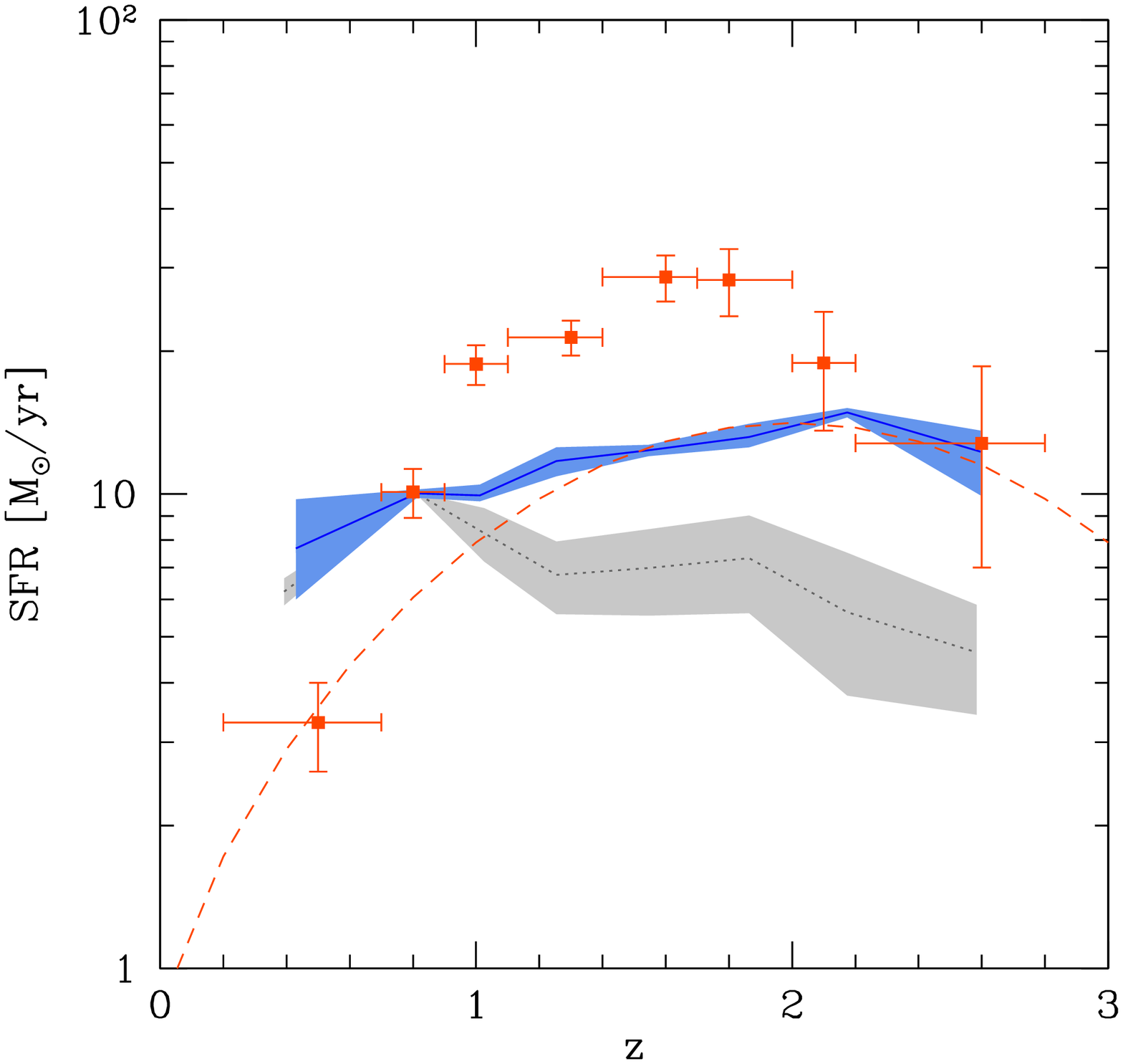}
\includegraphics[width=0.32\textwidth]{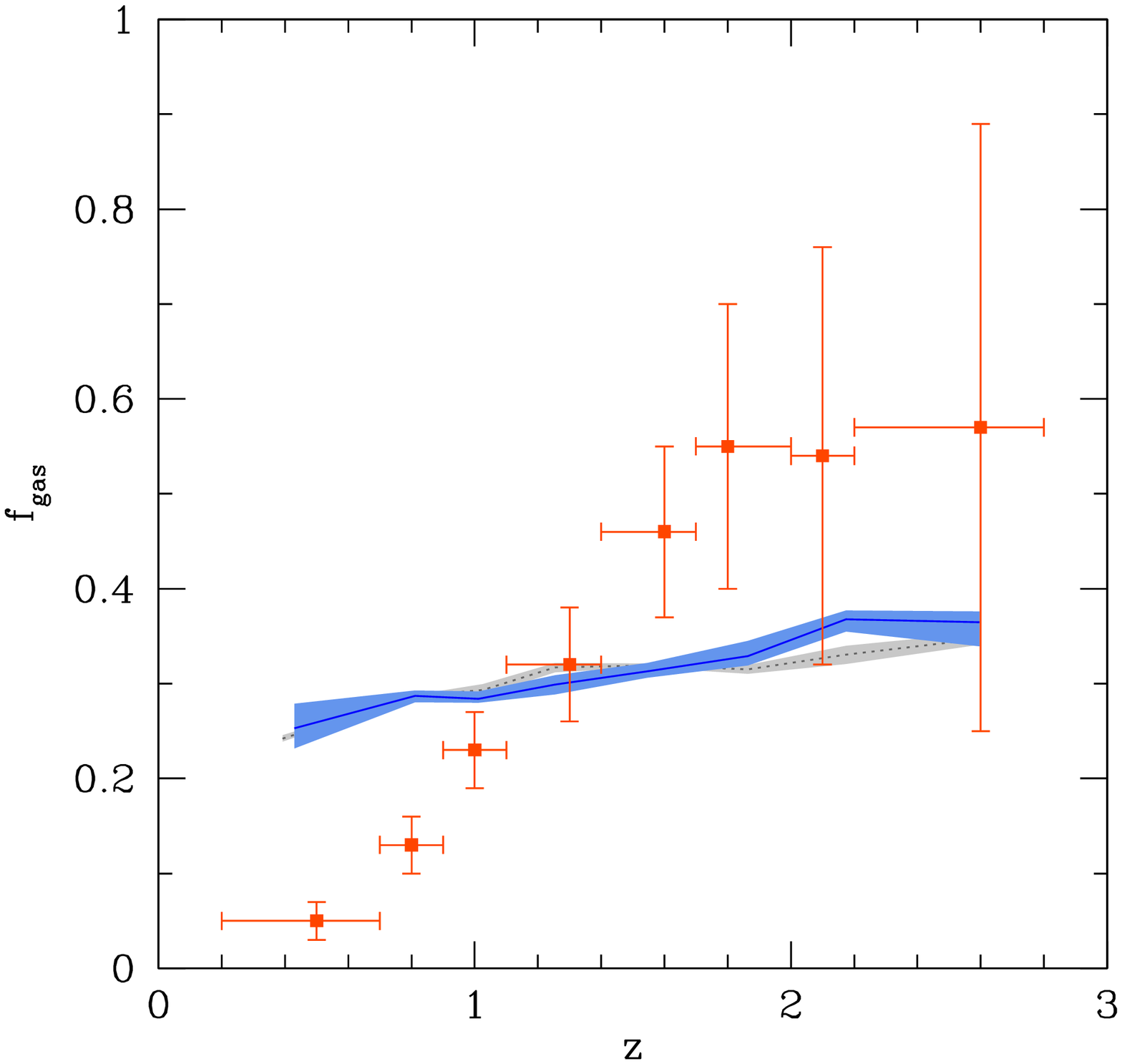}
%\vspace{-0.4cm}
\caption{Comparison between the stellar mass (left panel), SFR (middle panel) and gas fraction (right panel) of MW progenitors in the \texttt{GAMESH} simulation and observational data from \citet[red points]{2015ApJ...803...26P}.
 The blue solid lines show the average values among MW progenitors selected in each redshift bin  
following the same procedure adopted by \citet{2015ApJ...803...26P}. The grey dotted lines show the results
when the minimum mass adopted to select MW progenitors in each redshift bin is decreased by 1 dex (see text). 
The shaded regions represent the 1-$\sigma$ scatter around the mean.}
\label{fig:GAMESH-progMstar}
\end{figure*}

In addition, \citet{2014ApJ...786...54P} provide an estimate of the CGM metal mass as probed by low- and high-ionization (O~$\textsc{vi}$) species based on
COS-halos data and of the mass of dust in the circum-galactic medium (CGM) based on the reddening of background quasars \citep{2010MNRAS.405.1025M}. Taking the values from their
Table~5 and adding the metal mass in the hot X-ray emitting CGM gas\footnote{We compute this quantity from equation (24) in \citet{2014ApJ...786...54P} 
assuming a stellar mass of $[3-7]\times 10^{10} M_\odot$.}, we infer a total mass of heavy elements in the CGM of $1.7 \times 10^8 M_\odot$ with
minimum (maximum) values of $0.9\, (3.7)\times 10^8 M_\odot$. Hence, the total estimated mass of metals is found to be in the range 
$[1.85-8.4]\times 10^8 M_\odot$. Our model predicts a total mass of metals in the final MW halo to be $\sim 10^9 M_\odot$. However, at the MW mass
scale, the observations
probe the metal content within 150~kpc \citep{2014ApJ...786...54P}. If we assume the metals to follow the dark matter halo radial profile at these large radii, we find 
$M_{\rm Z}(r < 150~{\rm kpc})$ predicted by the simulation to be $6.9 \times 10^8 M_\odot$, in agreement with the observed value.
 
Hence, we conclude that having selected the two free parameters to be $\epsilon_{\ast} = 0.09$ and $\epsilon_{\rm w}= 0.0016$
the stellar mass and the mass of gas and metals in the ISM predicted by \texttt{GAMESH} for the MW-like halo at $z=0$ are consistent with the observations. A multiphase treatment of the gas and dust evolution in \texttt{GAMESH}, similar to what presented in DB14 and DB16, and the
relaxation of the IRA will be implemented in the future to address specific problems and are not expected to sensitively change the previous calibration.

In Figure~\ref{fig:GAMESH_CALIB_SFR} we show the total star formation rate (SFR) predicted by our model as function of the lookback time. 
Data points indicate observationally inferred estimates found in the literature, from the oldest estimates by \citet[SFR$\sim 5$~M$_{\odot}$ yr$^{-1}$, dark green cross]{1978A&A....66...65S}, \citet[SFR$\sim 4$~M$_{\odot}$ yr$^{-1}$, violet square]{1997ApJ...476..144M, 2006Natur.439...45D} to the newest, generally lower values around SFR $\sim 2$~M$_{\odot}$ yr$^{-1}$ by \citet[orange minus]{2010ApJ...710L..11R}, \citet[light-violet circle]{2015ApJ...806...96L}, \citet[red triangle]{2015A&A...580A.126K}.

As exhaustively discussed in \citet{2011AJ....142..197C}, the number of assumptions needed to derive the total SFR of the current Milky Way (for example in its structure, its stellar sample and stellar IMF) is so large that the resulting scatter can span one order of magnitude 
(see the azure diamond with the largest errorbars). For the MW-like halo at $z = 0$, \texttt{GAMESH} finds a SFR$ \sim 4.7$~M$_{\odot}$ yr$^{-1}$, a factor of two higher than recent estimates but still compatible with the data scatter. 
In the enclosed panel, we show the same SFR as a function of redshift $z$, to better visualize the evolution at high redshift.

While the global trend of the SFR predicted by \texttt{GAMESH} agrees with the one found in DB16, and the two SFRs show 
similar peak values ($\sim 15 - 20$~M$_{\odot}$ yr$^{-1}$), they peak at different redshits. 
The progressive, quasi parallel decline of the total SFR, results in final SFRs differing by  $\sim 1.5$~M$_{\odot}$ yr$^{-1}$. 
As argued for the gas and stellar mass behaviors, we ascribe these discrepancies to intrinsic differences in the merger tree definitions discussed above, MW mass assembly history, particularly at $z < 4$, and to the IRA, that naturally introduces an acceleration in the evolution. 

A comparison of \texttt{GAMESH} and its predictions with independent SAM models can be found in Appendix B. 

\section{Properties of Milky Way progenitors}

So far, we have investigated the global properties of the simulated halos, in the MW merger tree and in the LG. In this section, we discuss the SFR, the mass in stars, gas and metals predicted for MW progenitor systems at $0 < z < 4$ by the
\texttt{GAMESH} simulations and compare these with observations.

Recent studies have started to investigate the redshift evolution of progenitors of MW-like galaxies at $z = 0$, selecting
candidates from very deep near-IR surveys on the basis of their constant comoving density \citep{2013ApJ...771L..35V},
of their evolution on the galaxy star forming main sequence \citep{2013ApJ...778..115P}, or of multi-epoch abundance
matching techniques \citep{2015ApJ...803...26P}. 

Using a combined data sets based on the FourStar Galaxy Evolution
(ZFOURGE) survey, CANDELS {\it Hubble Space Telescope} (HST), 
{\it Spitzer}, and {\it Herschel}, \citet{2015ApJ...803...26P} derived 
photometric redshifts and stellar masses for MW progenitors and 
discuss their evolution with redshift. To compare
with their analysis, we have extracted MW progenitors from the 
 \texttt{GAMESH} simulation adopting a similar selection procedure. 
We first identify all the simulated systems in the same redshift bins
of the observations, and then we select those with a stellar mass
which falls within $\pm 0.25$ dex of the central stellar mass adopted
by \citet[see entries 1 and 2 in their Table 1]{2015ApJ...803...26P}.
The number of selected progenitors ranges between 41 (in the lowest
redshift bin, $0.2 < z < 0.7$) to 6 (in the highest redshift bin, $2.2 < z < 2.8$). 

Figure \ref{fig:GAMESH-progMstar} shows the resulting evolution of the average stellar mass 
(right panel), SFR (middle panel) and gas fraction (right panel), defined as 
$f_{\rm gas} = M_\ast/(M_{\rm gas}+M_{\ast})$. The blue solid lines show the model
predictions, with the shaded region representing the $1-\sigma$ scatter, and the 
red points are the \citet{2015ApJ...803...26P} data. To increase the
statistics of MW progenitors, particularly in the higher $z$ bins, we also show the 
model predictions when the mass selection is done within $\pm 1$ dex of the
central mass adopted by \citet[grey dotted line with shaded region]{2015ApJ...803...26P}.
Our simulation suggests that MW progenitors follow a stellar mass trend which is
in good agreement with the observations, particularly if the mass selection includes
a larger number of MW progenitor systems at $z > 1.5$.
In agreement with previous studies, we find that more than $90\%$
of the MW mass has been built since $z \sim 2.5$. 
However, the star formation rate and the gas fraction of the simulated galaxies
have a shallower evolution in the 3 Gyr period between $z = 2.5$ and $z = 1$ than
found by \citet{2015ApJ...803...26P}. In particular, the peak SFR of $\sim 10 \, M_{\odot}/{\rm yr}$ 
at $z \sim 1 - 2$ of the most massive MW progenitors is smaller than the value 
reported by  \citet{2015ApJ...803...26P}, and in closer agreement with the evolution
found by \citet[see the red dashed lines]{2013ApJ...771L..35V}. We find that the MW mass buildup can be fully explained by the 
SFRs of its progenitor systems, and does not require significant merging 
\citep{2013ApJ...771L..35V}. If star formation
dominated the formation of the MW galaxy, then its growth must
heavily depend on the evolution of cold gas and gas-accretion histories.
This is consistent with the results presented in Section 2.
In addition, by inverting the Kennicutt-Schmidt law, \citet{2015ApJ...803...26P} show that
the effective size and SFRs imply that the baryonic cold-gas 
fractions drop as galaxies evolve from high redshift to $z \sim 0$ (see the red 
data points in the right panel of  Figure~\ref{fig:GAMESH-progMstar}). The predicted
$f_{\rm gas}$ of the simulated sample show instead a rather flat trend and,
independently of the adopted selection criteria, the average SFR and gas fraction 
are larger than inferred by the observations below $z \sim 1$. 
The above mentioned discrepancies in the evolution of the average SFR and gas
fraction might be induced by the assumed Instantaneous Recycling Approximation, 
which affects the efficiency of mechanical and chemical feedback acting on the
evolution of individual galaxies, and by the lack of radiative transfer effects.
All these different feedback processes, indeed, can strongly affect the evolution
of the MW progenitors (e.g. \citealt{2010MNRAS.407L...1S, 2012MNRAS.421L..29S, 2015MNRAS.449.3137G}).

\begin{figure}
\centering
%\hskip -1.3truecm
\includegraphics[width=0.5\textwidth]{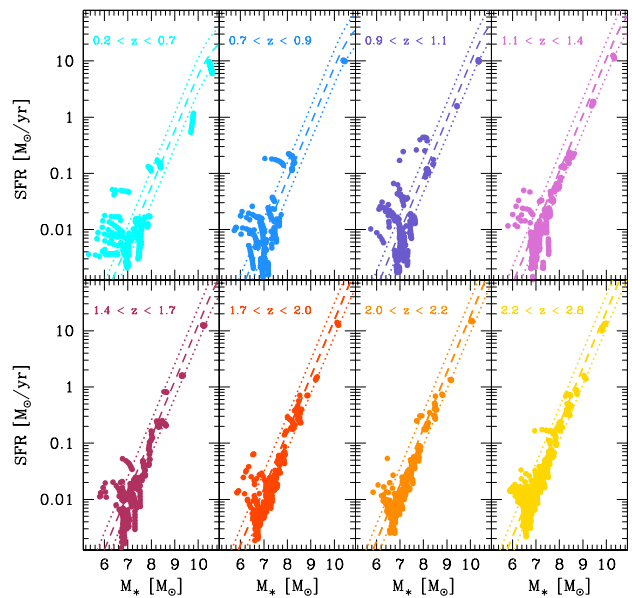}
%\vspace{-0.4cm}
\caption{The star formation rate as a function of stellar mass of all MW progenitors in different redshift bins, as indicated in the
legend. In each panel, the points represent the simulated systems and the dashed line shows the analytic fit to the galaxy main sequence 
at the central redshift of the bin, taken from \citet{2015A&A...575A..74S}.
The dotted lines are a factor of 2 above/below the fit.}
\label{fig:GAMESH-mainsequence}
\end{figure}

We have also checked the position of the simulated MW progenitors relative to the
galaxy main sequence of star formation. In Figure~\ref{fig:GAMESH-mainsequence}
we show the results using the same redshift bins adopted in Figure~\ref{fig:GAMESH-progMstar},
but without making any selection on the stellar mass. In each panel, the points represent all
the simulated systems, while the dashed line is the analytic fit to the observations,
taken from \citet[see their Eq.~9]{2015A&A...575A..74S} and computed at the central 
redshift of each bin. There is a large scatter in the SFR of the
smallest MW progenitors and most of the systems with $M_\ast < 10^8 M_\odot$ 
show SFRs that can vary by almost one order of magnitude.
While the galaxy main sequence can not be constrained by observations in this
regime, an increasing scatter towards low stellar masses has already been found in
hydro-dynamical simulations as a result of the
rising importance of stellar feedback \citep{2014MNRAS.445..581H}.  
Yet, the more massive among the MW progenitors
at each redshift lie within a factor of 2 of the galaxy main sequence 
(the region within the two dotted lines) all the way from $z \sim  2.5$ to $z \sim 0$. 

\begin{figure}
\centering
%\hskip -1.3truecm
\includegraphics[width=0.5\textwidth]{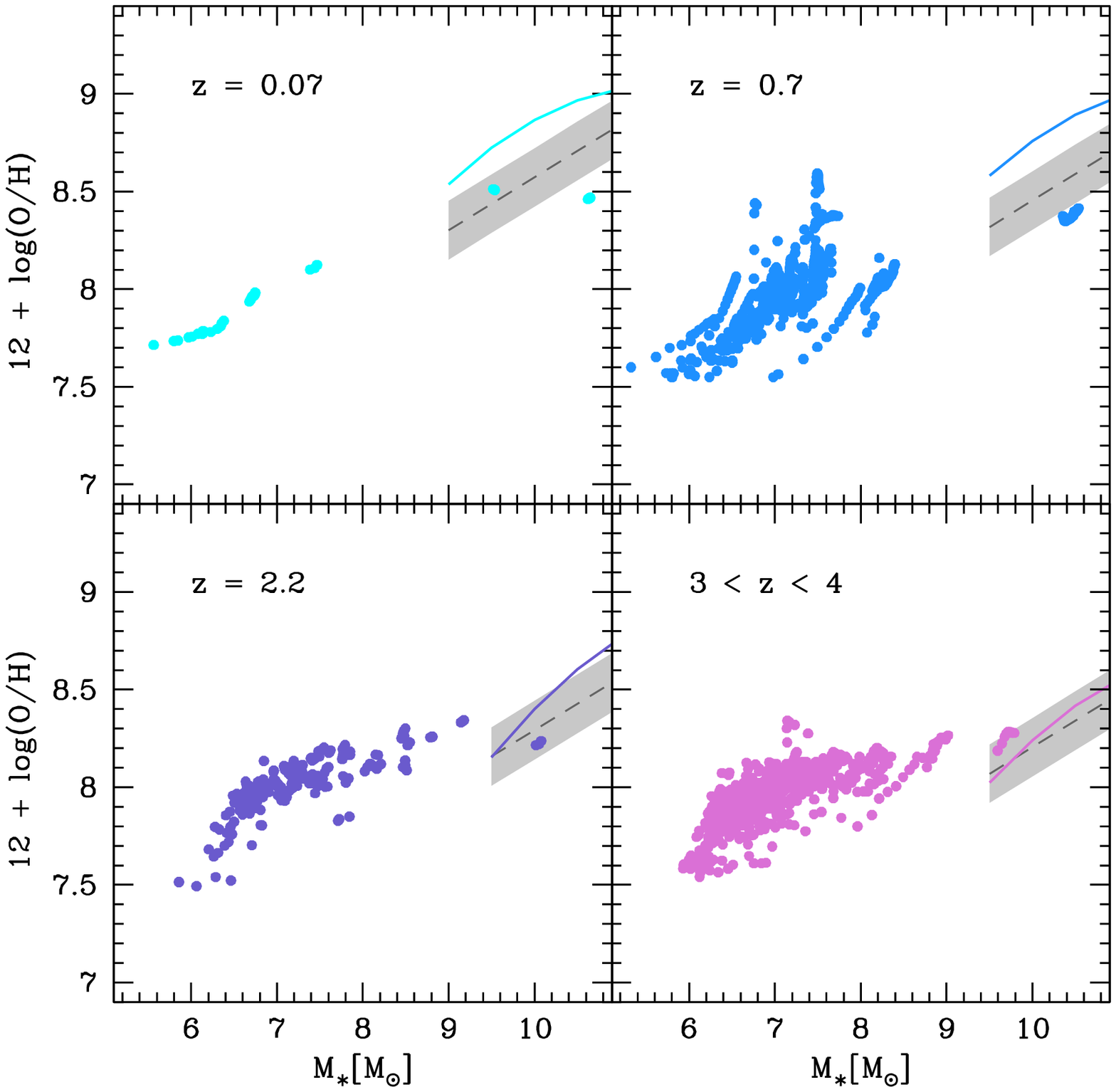}
%\vspace{-0.4cm}
\caption{The mass metallicity relation at different redshifts (see the legenda). The points
show the simulated MW progenitors and the solid lines represent the fit to the observed relations at
reported by \citet[at $z < 3$]{2008A&A...488..463M} and \citet[at $3 < z < 4$]{2009MNRAS.398.1915M}.
The dashed lines are the fit obtained by \citet{2016MNRAS.463.2002H} with the shaded region showing the
$\pm 0.15$ dex scatter. The tight relations followed by some of the points, particularly in the top right panel,
identify evolutionary tracks of  galaxies in the redshift range
encompassed by the observed samples.}
\label{fig:GAMESH-massmetallicity}
\end{figure}  

Finally, we compare the gas metallicity of the simulated MW progenitors with the observed
mass-metallicity relation (MZR) at different redshifts 
and with two (redshift independent) combinations of stellar mass, SFR and metallicity
known as the fundamental metallicity relation \citep[FMR]{2010MNRAS.408.2115M} and 
fundamental plane of metallicity \citep[FPZ]{2012MNRAS.427..906H,2016MNRAS.463.2002H}. 

\begin{figure*}
\centering
%\hskip -1.3truecm
\includegraphics[width=0.45\textwidth]{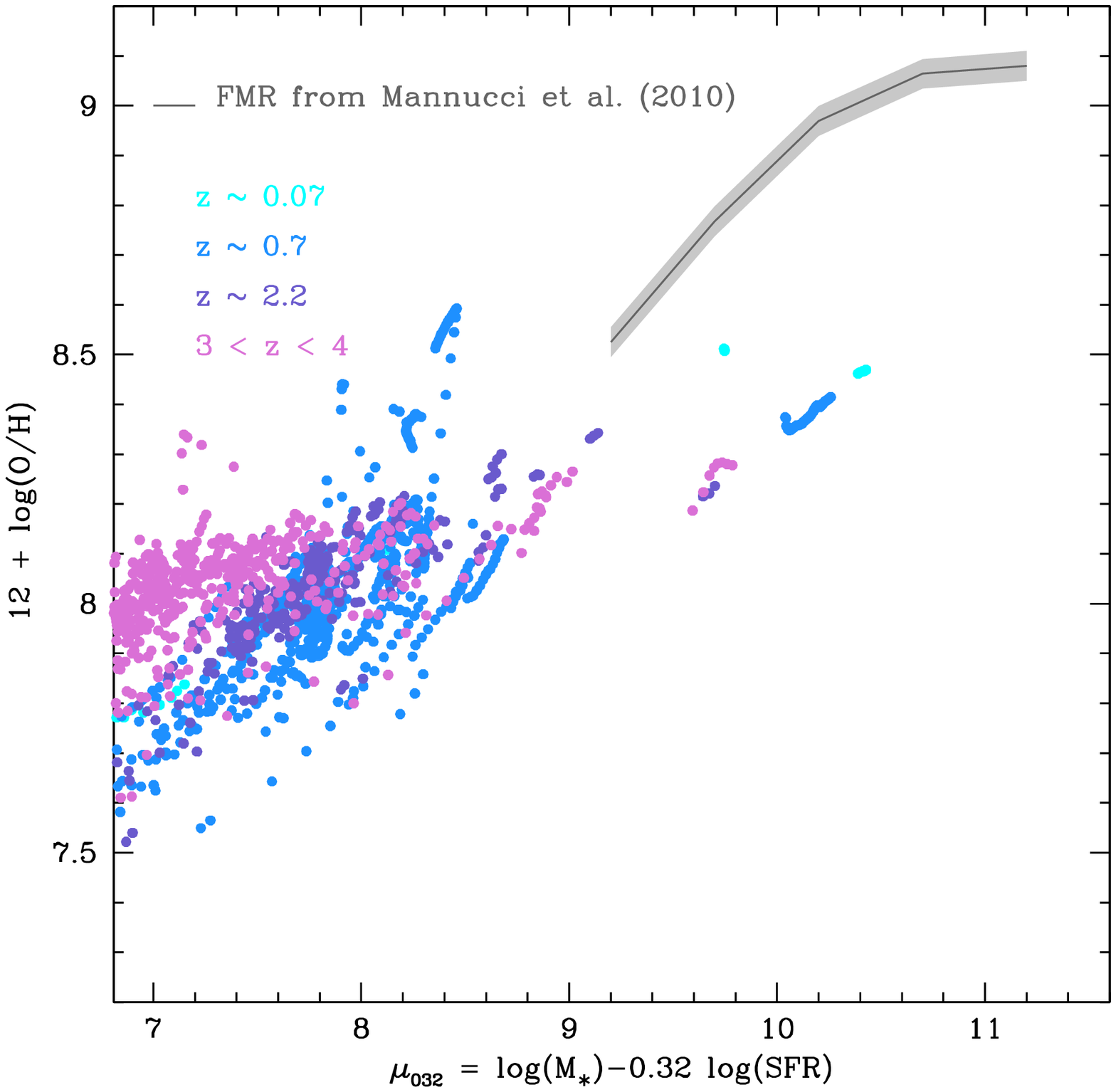}
\includegraphics[width=0.45\textwidth]{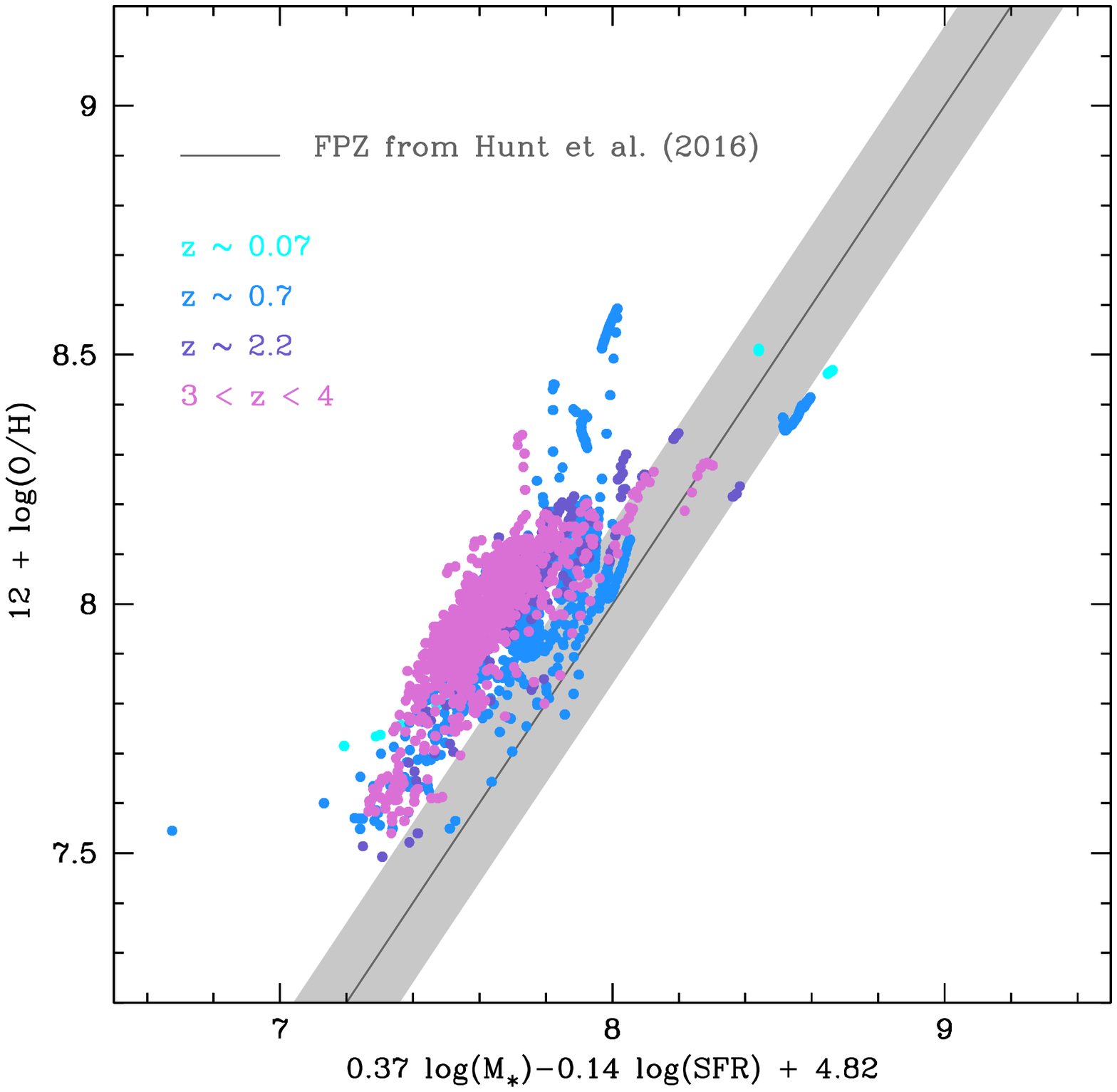}
%\vspace{-0.4cm}
\caption{Distribution of the MW progenitors relative to the fundamental metallicity relation
by \citet[left panel]{2010MNRAS.408.2115M} and to the fundamental plane of metallicity by \citet[right panel]{2016MNRAS.463.2002H}.
All the simulated systems at $\langle z \rangle = 0.07, 0.7, 2.2$ and at $3 < z < 4$
(using the same colour coding adopted in Figure~\ref{fig:GAMESH-massmetallicity}) are shown with data points. Dashed lines with
shaded regions show the observed fits (see text).}
\label{fig:GAMESH-FMRFPZ}
\end{figure*} 

The results are shown in Figs.~\ref{fig:GAMESH-massmetallicity} and \ref{fig:GAMESH-FMRFPZ}, respectively. 
The interstellar oxygen abundance has been computed assuming a solar oxygen-to-metal mass
ratio of $0.00674/0.0153 = 0.44$ \citep{2011SoPh..268..255C}, so that the solar metallicity 
corresponds to $12 + \rm{log(O/H)} = 8.759$.
The points represent all the simulated MW progenitors in the same redshift bins of 
\citet{2008A&A...488..463M} and \citet{2009MNRAS.398.1915M}, without any additional selection on the 
stellar mass or star formation rate. Instead, the solid lines show the fit to the data and
are drawn only in the mass range probed by the observations. 
It is clear that most of the simulated systems have stellar masses that are
outside this range, except for the few most massive MW progenitors with $M_{\ast} > 10^9 M_\odot$.
Similarly to Figure~\ref{fig:GAMESH-mainsequence}, the scatter in the MZR increases with
decreasing $M_\ast$, a result that appears to be consistent with deep spectroscopic 
observations which probe galaxies down to $M_\ast \sim 3 \times 10^7 M_\odot$ at $0.5 \le z \leq 0.7$
\citep{Guo2016}. 
At $3 < z < 4$, the most massive systems have metallicity slightly higher than those implied by the 
\citet{2009MNRAS.398.1915M} fit. However, at $z < 3$ the simulated systems fall systematically
below the fits by \citet{2008A&A...488..463M}. A better agreement is found with the fit to the
mass-metallicity relation proposed by \citet[and computed using their Eq.~2 at the average
redshift of each bin]{2016MNRAS.463.2002H}, 
shown as the dark grey dashed line, with the shaded region representing a dispersion of $\pm 0.15$ dex.

A similar result is found in Figure~\ref{fig:GAMESH-FMRFPZ}, where we show the position of the
simulated MW progenitors relative to the FMR (left panel) and FPZ (right panel). Here we have
reported systems selected in the same redshift bins as in Figure~\ref{fig:GAMESH-massmetallicity}.
As usual, smaller MW progenitors which populate the lower right side of the panels show a large
scatter at all redshifts. The most massive MW progenitors align along the FMR but with a -0.5 dex
metallicity offset. Conversely, their metallicity is within the scatter of the FPZ. 

We conclude that while the simulated systems may be slightly too metal-poor at high stellar masses 
and too metal-rich at lower stellar masses, the discrepancy with the MZR evolution by 
 \citet{2008A&A...488..463M} and with the FMR by  \citet{2010MNRAS.408.2115M} may be at
 least partly due to the different metallicity calibrations used by these authors, which may
 overestimate the observed metallicity by 0.3 dex \citep[see in particular their Section 4.4]{2016MNRAS.463.2002H}.
 (More details about discrepancies in metallicity calibrations can be found in \citealt{2008ApJ...681.1183K}.)

More interestingly, the distribution of the most massive MW progenitors is consistent with the FPZ
and aligned with the FMR. Since these redshift-independent scaling relations between metallicity, stellar
mass and star formation rate are believed to originate from the interplay between gas accretion, star formation and 
SN-driven outflows (see, among others, \citealt{2013MNRAS.430.2891D} and \citealt{2016MNRAS.463.2020H}), 
we conclude that the description of these physical processes in the \texttt{GAMESH} simulation 
leads to results consistent with observations at $0 < z < 4$ even in a simulated, biased region of the current Universe.

\begin{figure*}
\centering
  \includegraphics[trim=1.1cm 0cm 2.2cm 0cm, clip=true, width=0.49\textwidth]{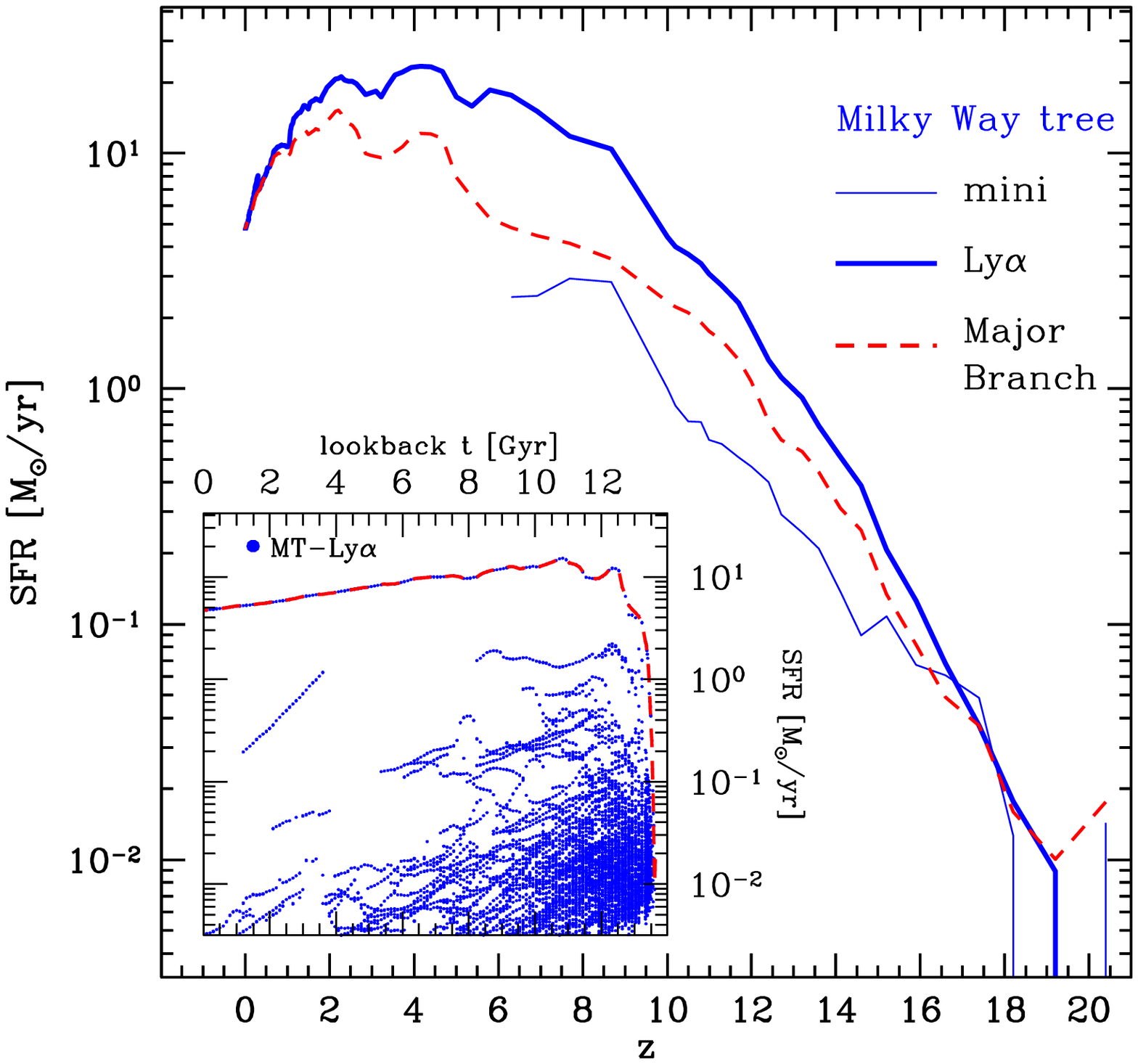}
  \includegraphics[trim=1.1cm 0cm 2.2cm 0cm, clip=true,  width=0.49\textwidth]{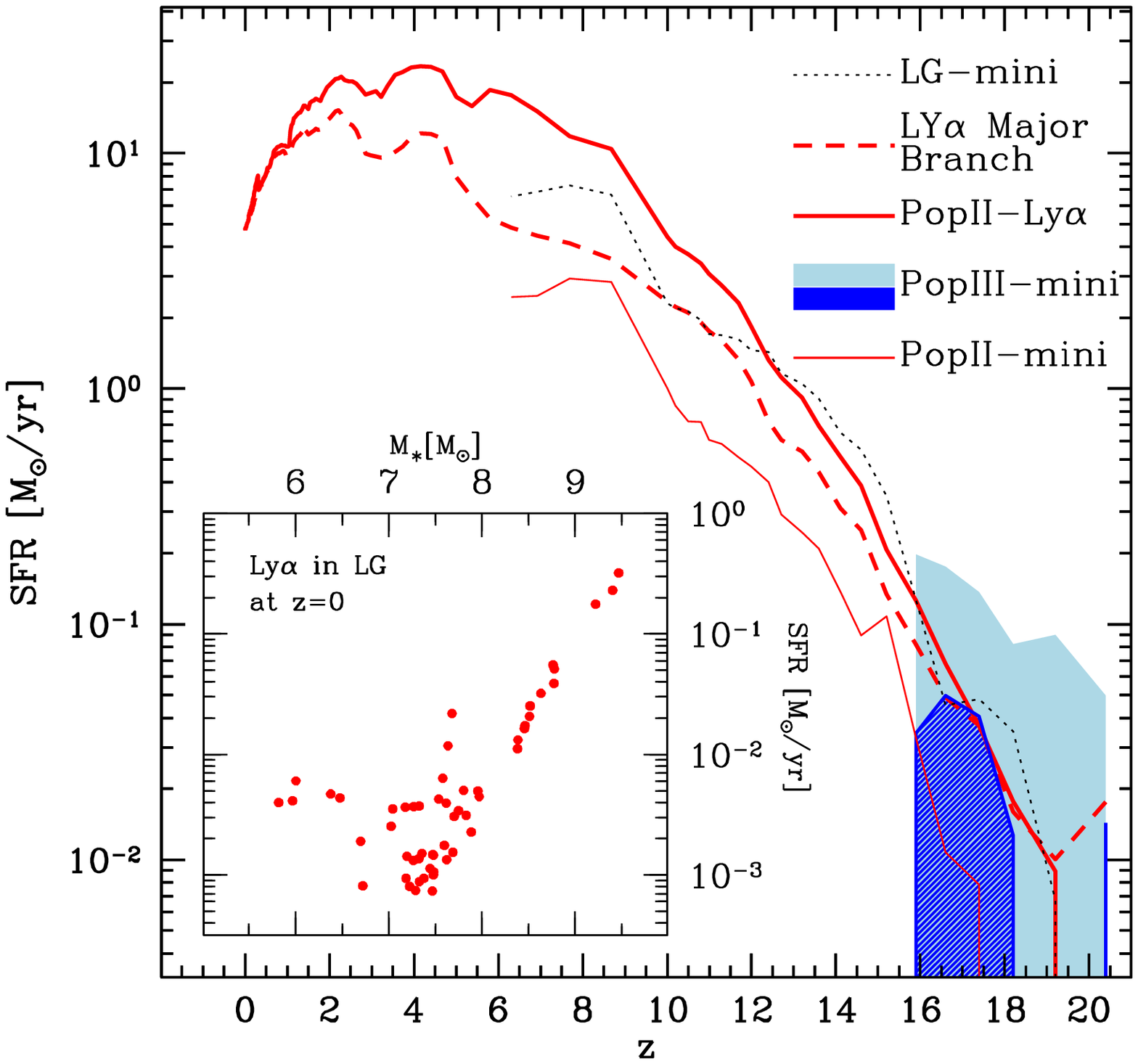}
\caption{Redshift evolution of the total SFR of galaxies hosted in mini- and Ly$\alpha$-cooling halos along the merger tree of the MW and in the LG. {\bf Left panel:}  SFR of mini- and Ly$\alpha$-cooling halos belonging to the MW merger tree (solid thin and thick blue lines, respectively).
The dashed red line shows the SFR along the major branch of the MW (the most massive halo at each redshift). 
The panel inset shows the SFR along the major branch (red dashed line)
as function of the lookback time $t$~[Gyr] and the SFR of all the Ly$\alpha$-cooling halos as blue points. 
{\bf Right panel:}  SFR of Pop\ II stars along the MW merger tree and hosted in mini-halos (solid thin red line) and in Ly$\alpha$-cooling halos
(solid thick red line). The dotted black line indicates the Pop\ II SFR history in all the mini-halos of the LG.  Pop\ III SFRs along the MW
merger tree and in the LG are indicated by shaded areas (blue and cyan, respectively). For comparison, we also show the SFR along the
MW MB (red dashed line). 
The panel inset shows the SFR versus stellar mass of all galaxies hosted by Ly$\alpha$-cooling halos in the LG at $z=0$.}
\label{fig:GAMESH-progMstar-MiniLyaHAlosSFR}
\end{figure*}  

\begin{figure}
\vspace{-1.0cm}
\subfloat{
  \includegraphics[clip, width=0.95\columnwidth]{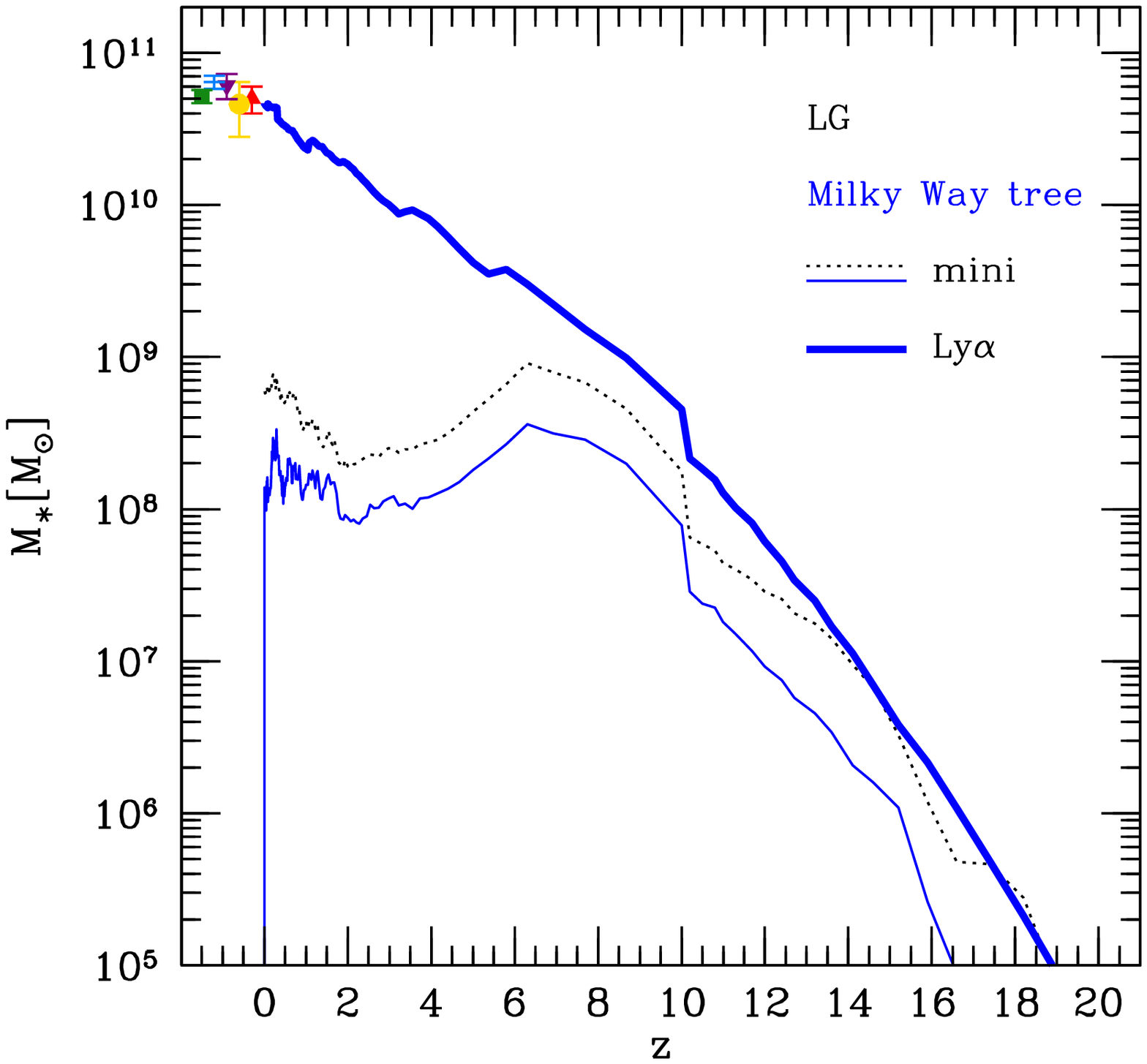}
}
\vspace{-2.1cm}
\subfloat{
  \includegraphics[clip, width=0.95\columnwidth]{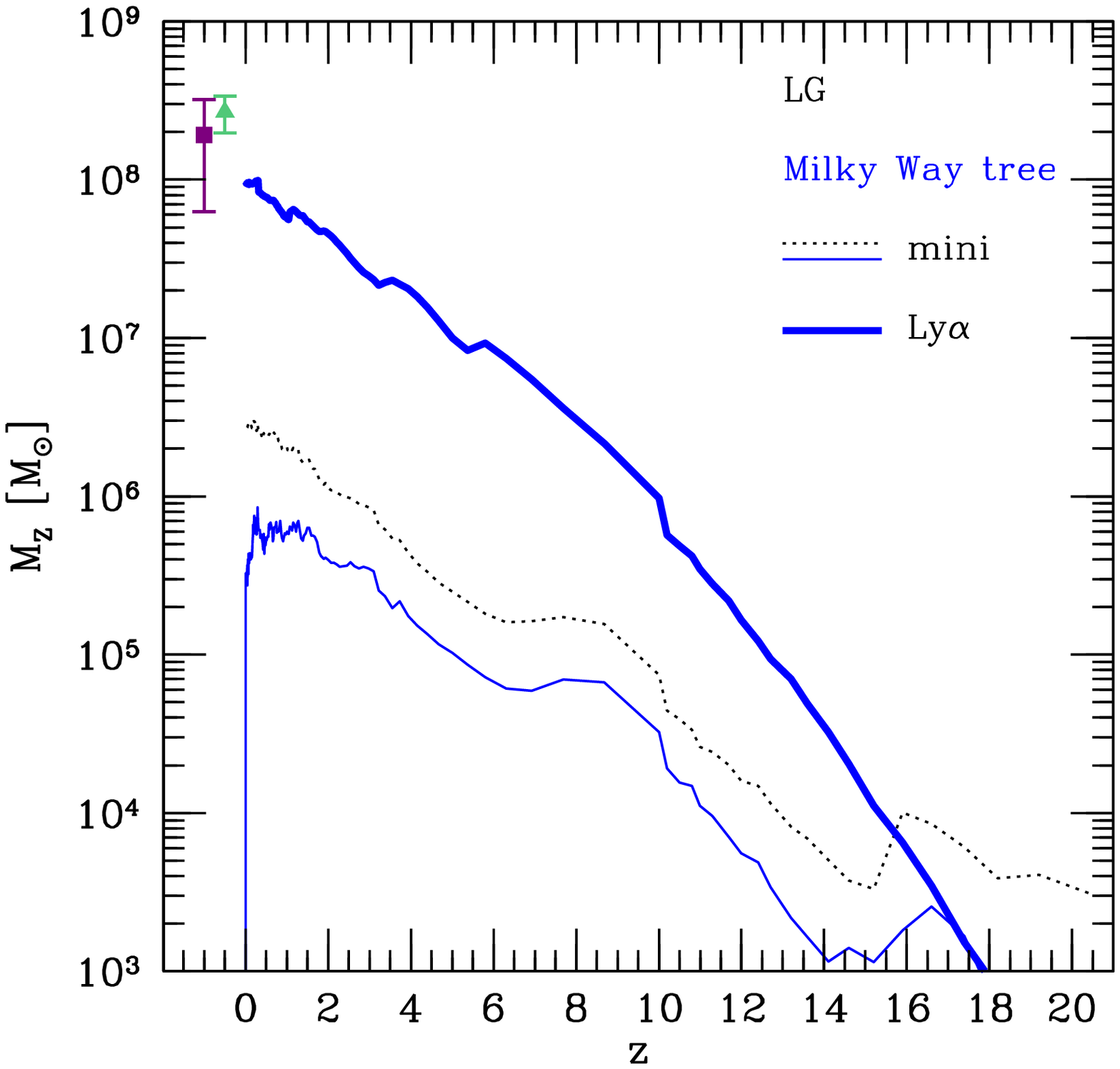}
}
\vspace{-1.2cm}
\caption{Contribution of galaxies hosted in mini- and Ly$\alpha$-cooling halos
to the total mass in stars ({\bf top panel}) and metals ({\bf bottom panel}) as a function of redshift.
MW progenitors hosted by mini-halos and Ly$\alpha$-cooling are shown by the solid thin and thick blue lines.
The dotted black lines show the systems hosted in mini-halos in the LG.}
\label{fig:GAMESH-progMstar-MiniLyaHAlosMSMZ}
\end{figure}  

\section{Evolution of galaxies hosted in mini- and Ly$\alpha$-cooling halos }
\label{sec:MiniLyaBaryons}

In this section we study the evolution of galaxy populations by assessing  their relative contribution 
to the total quantities discussed above. By following the standard classification of dark matter halos 
introduced in Section 2 (see Figure~\ref{fig:MWHaloPOPs}) we discuss the properties of the galaxies 
hosted in mini- and Ly$\alpha$-cooling halos. We first follow their SFR, M$_{*}$, M$_{Z}$ in time and then we
compare their final distribution at $z=0$.

In Figure~\ref{fig:GAMESH-progMstar-MiniLyaHAlosSFR} we show the redshift evolution of their SFR,
both in the MW merger tree and in the LG. 
In the left panel, the total SFR  of mini- and Ly$\alpha$-cooling halos found 
in the MW merger tree is shown by solid thin and thick blue lines, respectively. 
At all but the highest redshifts ($z>17$), the MW SFR  is dominated by galaxies hosted in 
Ly$\alpha$-cooling halos\footnote{The curves that show the contribution of mini-halos
are interrupted at $z = 6$ because we assume that SF is suppressed in mini-halos when $z \leq z_{\rm reio}$ (see Section 3).}. 
Despite the number of mini-halos is largely dominant, as shown in Figure~\ref{fig:MWHaloPOPs},  
their small contribution to the total SFR at all redshifts is due to (i) their low SF efficiency 
\citep{salvadori09, 2012MNRAS.421L..29S, 2015MNRAS.449.3137G} compared to Ly$\alpha$-cooling halos,
where the SFR is proportional to the halo gas content and controlled by the accretion history 
(see Figure~\ref{fig:MWHaloIGMAccretion}), (ii) their low gravitational potential, which implies an intrinsic 
inefficiency in accreting gas from the IGM.

Among Ly$\alpha$-cooling halos, a special role is played by the most massive halo which drives the
major branch of the MW. The SFR along the MB is shown by the red dashed line. The comparison
with the solid blue line shows that the MB dominates the SFR in the MW only at $z < 1$. This does not
come as a surprise, given that the MB contains half of the mass of the final MW halo at $z \lesssim 1.46$
and that at higher $z$ the MW mass assembly is dominated by a multiplicity of Ly$\alpha$-cooling systems,
which also dominate its SFR. This is clearly visible in the inset of the left panel of Figure~\ref{fig:GAMESH-progMstar-MiniLyaHAlosSFR},
where we show the SFR in the MB as a function of the lookback time (dashed red line)
and the SFR of individual Ly$\alpha$-cooling halos that are part of the MW merger tree but not collapsed in the MB (blue points).
The progressive disappearance of these points is a consequence of their accretion onto the MB. 
We note that the flat evolution of the SFR in the MB in the past 8 Gyr is in line with findings of independent  
models \citep{2014MNRAS.445..970D}.

In the right panel of Figure~\ref{fig:GAMESH-progMstar-MiniLyaHAlosSFR}, we investigate the relative 
contribution of Pop~II and Pop~III stars to the SF history in the merger tree of the MW and in the LG.
In the very early evolution ($z >16$), the SFR is dominated by Pop~III stars in both the LG and along the MW merger tree
(represented by the cyan and blue shaded areas, respectively). Due to the effect of metal enrichment, their formation
is mostly confined to the first star-forming mini-halos and the sharp drop in their SFR is driven 
by  the increase of the average metallicity of the IGM above $Z_{\rm crit}=10^{-4} Z_{\odot}$ at $z < 16$.
We note that the Pop~III SFR is larger in the LG than along the MW merger tree, as a result of the larger statistics of mini-halos. This suggests that traces of early Pop~III star formation are not confined to the MW and its satellites but may be found in external galaxies of the LG, although their detectability seems to remain beyond the capabilities of the new generation of telescopes, such as the JWST\footnote{{ Due to the intrinsic differences in the dynamical configuration of halos in our LG with respect to the observed one, this information should be valuated only in a statistical sense.} }. Finally note how the sharp transition between Pop\ III and Pop\ II stars is certainly affected by the absence in the current study of inhomogeneous RT and metal mixing in the LG, as predicted by independent models \citep{2003ApJ...589...35S, 2014MNRAS.437L..26S, 2016arXiv161100025S}. This point will be deeply investigated in the companion work of this paper including full RT and particle tagging (Graziani et al., in prep.).

The remaining redshift evolution is dominated by Pop~II stars formed in mini- and Ly$\alpha$-cooling halos (solid thin and thick red lines, 
respectively).
We also show the contribution of Pop~II star formation in mini-halos belonging to the LG  (black dotted line). It is interesting to
note that this component is comparable to the total SFR along the merger tree of the MW and it is larger than the SFR on the MB 
down to $z \approx z_{\rm reio}$, below which it is suppressed by reionization. Although our simplified description of radiative feedback 
may overestimate star formation in small objects prior to reionization, this comparison shows that 
they provide an important source of ionizing photons within the LG (\citealt{2014MNRAS.437L..26S}, Graziani et al. in prep).

The star forming systems found in the LG  at $z=0$  are shown in the figure inset, where their SFR is plotted as function of the stellar 
mass\footnote{Note that the MW galaxy is excluded.}. Their distribution follows the galaxy main sequence presented in Figure~\ref{fig:GAMESH-mainsequence}. 
The three most massive SF galaxies, with $M_{*} >10^9  M_{\odot}$, have $0.1<$SFR$<1 M_{\odot}$~yr$^{-1}$,
similar to the values typically found in big objects observed in the LG.
%%, i.e. M31 (SFR$\, \approx 0.25 \, M_{\odot}$~yr$^{-1}$, \citealt{2013ApJ...769...55F}). 
Similarly, galaxies with lower stellar mass, $10^8 <M_{*}/ M_{\odot} <10^9$ have  $10^{-2}<$ SFR $<10^{-1}$~$M_{\odot}$~yr$^{-1}$,
as observed in the Large (LMC) and Small Magellanic Clouds (SMC) which have $M_\ast \approx10^9$~$M_{\odot}$, $0.25 < $ SFR $<0.63 $~$M_{\odot}$~yr$^{-1}$  (LMC) and $M \approx 3 \times 10^8$~$M_{\odot}$, $0.016 < $ SFR $<0.039 $~$M_{\odot}$~yr$^{-1}$  (SMC, \citealt{Skibba2012}). Our simulation also finds a third 
population of galaxies with $M_{*} <10^8 \, M_{\odot}$ hosted in small Ly$\alpha$-cooling halos with SFR$<10^{-3}$~$M_{\odot}$~yr$^{-1}$,
within the detection limits of recent dwarf galaxy surveys, such as the DGS \citep{2013PASP..125..600M,2014PASP..126.1079M}.

In the two panels of Figure~\ref{fig:GAMESH-progMstar-MiniLyaHAlosMSMZ} we show
the redshift evolution of the stellar (top panel) and metal (bottom panel) mass , separating
the contribution of MW progenitors hosted in mini- and Ly$\alpha$-cooling halos (solid thin and thick blue lines, respectively),
and of all the mini-halos found in the LG (dotted black lines).  As expected, progenitors in Ly$\alpha$-cooling halos dominate the evolution
and the total stellar and metal mass in mini-halos is orders of magnitude smaller, even when all the mini-halos in the LG are considered
(with the exception of the metal mass at $z > 16$, due to efficient metal enrichment by Pop~III stars).
The evolution of the stellar mass of both Ly$\alpha$-cooling and mini-halos is in good agreement with the
results of \citet{2014MNRAS.437L..26S} that match the observed properties of MW dwarf galaxies using a more 
sophisticated version of the semi-analytic model in which the IRA approximation is relaxed and 
the inhomogeneous metal enrichment and reionization of the MW environment is accounted for 
by using analytic prescriptions.

At $z < 6$, the stellar mass in mini-halos decreases as a result of the combined effect of radiative feedback following reionization
and assembly of mini-halos into bigger structures (see also Figure~\ref{fig:MWHaloPOPs}).

Below $z < 2$, the oscillating behavior in the thin blue lines are due to continuous mass exchange
between small progenitors (both mini and small Ly$\alpha$-cooling halos) orbiting around the MB, 
before the final merging, which causes the sharp drop at $z=0$\footnote{By definition, the MW merger tree
contains only the final MW host halo at $z=0$.} A series of mass exchange events through dynamical interactions or 
destructive mergers involving small Ly$\alpha$-cooling and mini-halos is at the origin of the final, spiky mass increase  
in the dotted black lines, referring to small dwarf galaxies surviving in the LG at $z=0$. By checking the merger trees of these small objects surviving in the Local Universe, we verified that the increase in metallicity of the mini-halo population is 
mainly ascribable to a series of contamination events from highly enriched baryon masses tidally dragged by mini-halos orbiting Ly$\alpha$-cooling halos. An increasing number of small Ly$\alpha$-cooling halos transitioning the mini-halos population after mass loss by tidal interactions is also found in the lowest redshift evolution. A deeper understanding of this interesting interplay between systems dominated by interacting dynamics and sharing tracers of chemical evolution, certainly requires the adoption of a full particle scheme and it is the deferred to a future work.     

\section{Conclusions}

In this paper we explored the properties of Milky Way progenitors by simulating the Galaxy formation 
process with the \texttt{GAMESH} pipeline. To guarantee a good statistics of halo progenitors, \texttt{GAMESH} 
adopted a new DM simulation providing a well mass resolved cosmological box containing a central Milky 
Way-like halo and a significant number of smaller objects and MW satellites. 
This new simulation allows us to draw the following, main conclusions:

\begin{itemize}  

\item The new DM simulation has the adequate mass resolution to guarantee a reliable Milky Way-type DM halo
      whose mass and dynamical properties are in agreement with a number of independent simulations and 
      observational estimates. The Local Group surrounding it shows a plethora of intermediate 
      mass Lyman $\alpha$-cooling halos and a vast number of mini-halos, useful to study both the global 
      accretion process and the effects of mergers and tidal stripping.      
      The time resolution of the new simulation has proven to be adequate to reproduce the major events of
      the accretion history of the MW halo, in agreement with similar trends found by independent simulations. 
      
\item Once processed by \texttt{GAMESH} and after tuning the star formation and galactic 
      wind efficiency, the stellar, gas and metal mass at $z=0$ are consistent with Milky Way observations. 
      We predict a final SFR for the resulting Milky Way system
      a factor of 2 higher than recent simulations but still in agreement with the many observational uncertainties.  

\item A particle-by-particle reconstruction of the Milky Way merger tree allows us to follow the redshift evolution of 
      MW progenitor galaxies and to predict their baryonic properties. 
      The simulated progenitor galaxies follow a stellar mass trend in good agreement with observations
      targeting ``plausible'' MW progenitors \citep{2013ApJ...771L..35V, 2015ApJ...803...26P}. Our simulation 
      suggests that more than 90\% of the MW mass has been built since $z \sim 2.5$. 
      However, the star formation rate and the gas fraction of the simulated galaxies have a shallower evolution 
      between $z = 2.5$ and $z = 1$ than 
      found by \citet{2015ApJ...803...26P}. While the MW mass build-up can be fully explained by the SFRs of 
      its progenitor systems, and does not require significant merging, we will re-evaluate all the discrepancies 
      found with observations in future model implementations relaxing 
      the IRA approximation and accounting for detailed radiative feedback.

\item The most massive among the simulated MW progenitors lie within a factor of 2 of the galaxy main sequence 
      all the way from $z \sim 2.5$ to $z \sim 0$. The predicted SFRs show an increasing scatter towards low stellar 
      mass systems due to the rising importance of feedback effects. Similar results are found when comparing the 
      distribution of the simulated galaxies with the observed mass-metallicity relation, fundamental metallicity
      relation and fundamental plane of metallicity at $0 < z < 4$  
      \citep{2010MNRAS.408.2115M, 2012MNRAS.427..906H, 2016MNRAS.463.2002H}.
      Since these scaling relations are believed to originate from the interplay between gas accretion, 
      star formation and SN-driven outflows, we conclude that the description of these physical 
      processes obtained by \texttt{GAMESH} leads to results consistent with observations. 
      
\item At all but the highest redshifts, the SFR of the Milky Way is dominated by a multiplicity
      of galaxies hosted in Lyman $\alpha$-cooling halos, hosting Pop II stars. These systems are progressively
      accreted by the major branch of the MW merger tree, which provides the dominant contribution to the
      SFR at $z < 1$. The cumulative contribution of star forming mini-halos in the Local Group is comparable to
      the SFR along the MW merger tree at $z > 6$, indicating that these systems provide an important source
      of ionizing photons.
      
\item Due to efficient metal enrichment, Pop~III stars are confined to form in the smallest
      mini-halos at $z > 16$, and their formation rate is larger in the Local Group than along the MW merger tree.
      This suggests that traces of Pop~III star formation are not confined to the MW and its satellites but may be
      found in external galaxies of the Local Group, although their detection may be challenging even for the 
      next generation of telescopes.

\item We find that a large number of mini-halos having old stellar populations are dragged into the MW or can survive in the local Universe.  
         However, due to the effect of radiative feedback, mini-halos collapsing at $z < z _{\rm reio}$ remain instead dark because they never 
         experienced star formation.
     
\item The low redshift evolution of all halos, when followed in stellar mass and metal mass, shows the importance of     	     dynamical effects acting onto progenitors which are being accreted on the major branch of the MW. Events of late 	     mergers, tidal stripping and halo disruptions are found to be relevant in redistributing baryonic properties among 
     halo families, and also prove the capability of \texttt{GAMESH} in tracking the statistical relevance of dynamical    		     effects without accounting for a detailed gas dynamics treatment.    
     
\end{itemize}

The work that we have presented represents a promising starting point for a more detailed analysis 
based on sophisticated simulations having a proper treatment of the radiative transfer and 
inhomogeneous metal enrichment with \texttt{GAMESH}. 

\section*{Acknowledgments}
The authors would like to thank Leslie Hunt, Giulia Rodighiero and the anonymous referee for their very constructive comments. 
LG thanks Michele Ginolfi for his support with observational data.
SS was supported by the European Commission through an individual Marie-Skodolowska-Curie Fellowship, project: PRIMORDIAL, 700907.
The cosmological simulation and their test simulations for this paper were ran on the UCL facility Grace, and the DiRAC Facilities (through the COSMOS and MSSL-Astro consortium) jointly funded by the UK \textquotesingle s Science \& Technology Facilities Council and the Large Facilities Capital Fund of BIS.
We also acknowledge PRACE 
\footnote{http://www.prace-ri.eu/} for awarding us access to the CEA HPC facility "CURIE@GENCI"
\footnote{http://www-hpc.cea.fr/en/complexe/tgcc-curie.htm} with the Type B project: High 
Performance release of the GAMESH pipeline.

The research leading to these results has received funding from the European 
Research Council under the European Union's Seventh Framework Programme 
(FP/2007-2013) / ERC Grant Agreement n. 306476.

\bibliographystyle{mn2e}
\bibliography{GAMESH2}

\begin{thebibliography}{}
\makeatletter
\relax
\def\mn@urlcharsother{\let\do\@makeother \do\$\do\&\do\#\do\^\do\_\do\%\do\~}
\def\mn@doi{\begingroup\mn@urlcharsother \@ifnextchar [ {\mn@doi@}
  {\mn@doi@[]}}
\def\mn@doi@[#1]#2{\def\@tempa{#1}\ifx\@tempa\@empty \href
  {http://dx.doi.org/#2} {doi:#2}\else \href {http://dx.doi.org/#2} {#1}\fi
  \endgroup}
\def\mn@eprint#1#2{\mn@eprint@#1:#2::\@nil}
\def\mn@eprint@arXiv#1{\href {http://arxiv.org/abs/#1} {{\tt arXiv:#1}}}
\def\mn@eprint@dblp#1{\href {http://dblp.uni-trier.de/rec/bibtex/#1.xml}
  {dblp:#1}}
\def\mn@eprint@#1:#2:#3:#4\@nil{\def\@tempa {#1}\def\@tempb {#2}\def\@tempc
  {#3}\ifx \@tempc \@empty \let \@tempc \@tempb \let \@tempb \@tempa \fi \ifx
  \@tempb \@empty \def\@tempb {arXiv}\fi \@ifundefined
  {mn@eprint@\@tempb}{\@tempb:\@tempc}{\expandafter \expandafter \csname
  mn@eprint@\@tempb\endcsname \expandafter{\@tempc}}}

\bibitem[\protect\citeauthoryear{{Abazajian} et~al.,}{{Abazajian}
  et~al.}{2003}]{2003AJ....126.2081A}
{Abazajian} K.,  et~al., 2003, \mn@doi [AJ] {10.1086/378165}, \href
  {http://adsabs.harvard.edu/abs/2003AJ....126.2081A} {126, 2081}

\bibitem[\protect\citeauthoryear{{Alam} et~al.,}{{Alam}
  et~al.}{2015}]{2015ApJS..219...12A}
{Alam} S.,  et~al., 2015, \mn@doi [ApJS] {10.1088/0067-0049/219/1/12}, \href
  {http://adsabs.harvard.edu/abs/2015ApJS..219...12A} {219, 12}

\bibitem[\protect\citeauthoryear{{Angulo}, {Springel}, {White}, {Jenkins},
  {Baugh}  \& {Frenk}}{{Angulo} et~al.}{2012}]{2012MNRAS.426.2046A}
{Angulo} R.~E.,  {Springel} V.,  {White} S.~D.~M.,  {Jenkins} A.,  {Baugh}
  C.~M.,   {Frenk} C.~S.,  2012, \mn@doi [MNRAS]
  {10.1111/j.1365-2966.2012.21830.x}, \href
  {http://adsabs.harvard.edu/abs/2012MNRAS.426.2046A} {426, 2046}

\bibitem[\protect\citeauthoryear{{Behroozi}, {Wechsler}  \&
  {Conroy}}{{Behroozi} et~al.}{2013}]{2013ApJ...770...57B}
{Behroozi} P.~S.,  {Wechsler} R.~H.,   {Conroy} C.,  2013, \mn@doi [ApJ]
  {10.1088/0004-637X/770/1/57}, \href
  {http://adsabs.harvard.edu/abs/2013ApJ...770...57B} {770, 57}

\bibitem[\protect\citeauthoryear{{Bland-Hawthorn} \&
  {Gerhard}}{{Bland-Hawthorn} \& {Gerhard}}{2016}]{2016ARA&A..54..529B}
{Bland-Hawthorn} J.,  {Gerhard} O.,  2016, \mn@doi [ARAA]
  {10.1146/annurev-astro-081915-023441}, \href
  {http://adsabs.harvard.edu/abs/2016ARA%26A..54..529B} {54, 529}

\bibitem[\protect\citeauthoryear{{Bland-Hawthorn}, {Sutherland}  \&
  {Webster}}{{Bland-Hawthorn} et~al.}{2015}]{2015ApJ...807..154B}
{Bland-Hawthorn} J.,  {Sutherland} R.,   {Webster} D.,  2015, \mn@doi [ApJ]
  {10.1088/0004-637X/807/2/154}, \href
  {http://adsabs.harvard.edu/abs/2015ApJ...807..154B} {807, 154}

\bibitem[\protect\citeauthoryear{{Blanton} et~al.,}{{Blanton}
  et~al.}{2017}]{2017arXiv170300052B}
{Blanton} M.~R.,  et~al., 2017, preprint, \href
  {http://adsabs.harvard.edu/abs/2017arXiv170300052B} {} (\mn@eprint {arXiv}
  {1703.00052})

\bibitem[\protect\citeauthoryear{{Bouwens} et~al.,}{{Bouwens}
  et~al.}{2016}]{2016ApJ...830...67B}
{Bouwens} R.~J.,  et~al., 2016, ApJ, 830, 67

\bibitem[\protect\citeauthoryear{{Bovill} \& {Ricotti}}{{Bovill} \&
  {Ricotti}}{2009}]{2009ApJ...693.1859B}
{Bovill} M.~S.,  {Ricotti} M.,  2009, ApJ, 693, 1859

\bibitem[\protect\citeauthoryear{{Bovy} \& {Rix}}{{Bovy} \&
  {Rix}}{2013}]{2013ApJ...779..115B}
{Bovy} J.,  {Rix} H.-W.,  2013, \mn@doi [ApJ] {10.1088/0004-637X/779/2/115},
  \href {http://adsabs.harvard.edu/abs/2013ApJ...779..115B} {779, 115}

\bibitem[\protect\citeauthoryear{{Boylan-Kolchin}, {Springel}, {White},
  {Jenkins}  \& {Lemson}}{{Boylan-Kolchin} et~al.}{2009}]{2009MNRAS.398.1150B}
{Boylan-Kolchin} M.,  {Springel} V.,  {White} S.~D.~M.,  {Jenkins} A.,
  {Lemson} G.,  2009, MNRAS, 398, 1150

\bibitem[\protect\citeauthoryear{{Bromm} \& {Yoshida}}{{Bromm} \&
  {Yoshida}}{2011}]{2011ARA&A..49..373B}
{Bromm} V.,  {Yoshida} N.,  2011, \mn@doi [ARAA]
  {10.1146/annurev-astro-081710-102608}, \href
  {http://adsabs.harvard.edu/abs/2011ARA%26A..49..373B} {49, 373}

\bibitem[\protect\citeauthoryear{{Brook}, {Kawata}, {Scannapieco}, {Martel}  \&
  {Gibson}}{{Brook} et~al.}{2007}]{2007ApJ...661...10B}
{Brook} C.~B.,  {Kawata} D.,  {Scannapieco} E.,  {Martel} H.,   {Gibson} B.~K.,
   2007, \mn@doi [ApJ] {10.1086/511514}, \href
  {http://adsabs.harvard.edu/abs/2007ApJ...661...10B} {661, 10}

\bibitem[\protect\citeauthoryear{{Brown} et~al.,}{{Brown}
  et~al.}{2014}]{2014ApJ...796...91B}
{Brown} T.~M.,  et~al., 2014, ApJ, 796, 91

\bibitem[\protect\citeauthoryear{{Caffau}, {Ludwig}, {Steffen}, {Freytag}  \&
  {Bonifacio}}{{Caffau} et~al.}{2011}]{2011SoPh..268..255C}
{Caffau} E.,  {Ludwig} H.-G.,  {Steffen} M.,  {Freytag} B.,   {Bonifacio} P.,
  2011, Solar Phys, 268, 255

\bibitem[\protect\citeauthoryear{{Carlesi} et~al.,}{{Carlesi}
  et~al.}{2016}]{2016MNRAS.458..900C}
{Carlesi} E.,  et~al., 2016, \mn@doi [MNRAS] {10.1093/mnras/stw357}, \href
  {http://adsabs.harvard.edu/abs/2016MNRAS.458..900C} {458, 900}

\bibitem[\protect\citeauthoryear{{Chomiuk} \& {Povich}}{{Chomiuk} \&
  {Povich}}{2011}]{2011AJ....142..197C}
{Chomiuk} L.,  {Povich} M.~S.,  2011, \mn@doi [AJ]
  {10.1088/0004-6256/142/6/197}, \href
  {http://adsabs.harvard.edu/abs/2011AJ....142..197C} {142, 197}

\bibitem[\protect\citeauthoryear{{Ciardi} \& {Ferrara}}{{Ciardi} \&
  {Ferrara}}{2005}]{2005SSRv..116..625C}
{Ciardi} B.,  {Ferrara} A.,  2005, \mn@doi [SSR] {10.1007/s11214-005-3592-0},
  \href {http://adsabs.harvard.edu/abs/2005SSRv..116..625C} {116, 625}

\bibitem[\protect\citeauthoryear{{Ciardi}, {Ferrara}, {Marri}  \&
  {Raimondo}}{{Ciardi} et~al.}{2001}]{2001MNRAS.324..381C}
{Ciardi} B.,  {Ferrara} A.,  {Marri} S.,   {Raimondo} G.,  2001, \mn@doi
  [MNRAS] {10.1046/j.1365-8711.2001.04316.x}, \href
  {http://adsabs.harvard.edu/abs/2001MNRAS.324..381C} {324, 381}

\bibitem[\protect\citeauthoryear{{Cole}, {Aragon-Salamanca}, {Frenk}, {Navarro}
   \& {Zepf}}{{Cole} et~al.}{1994}]{1994MNRAS.271..781C}
{Cole} S.,  {Aragon-Salamanca} A.,  {Frenk} C.~S.,  {Navarro} J.~F.,   {Zepf}
  S.~E.,  1994, \mn@doi [MNRAS] {10.1093/mnras/271.4.781}, \href
  {http://adsabs.harvard.edu/abs/1994MNRAS.271..781C} {271, 781}

\bibitem[\protect\citeauthoryear{{Cole}, {Lacey}, {Baugh}  \& {Frenk}}{{Cole}
  et~al.}{2000}]{2000MNRAS.319..168C}
{Cole} S.,  {Lacey} C.~G.,  {Baugh} C.~M.,   {Frenk} C.~S.,  2000, \mn@doi
  [MNRAS] {10.1046/j.1365-8711.2000.03879.x}, \href
  {http://adsabs.harvard.edu/abs/2000MNRAS.319..168C} {319, 168}

\bibitem[\protect\citeauthoryear{{Colin}, {Avila-Reese}, {Roca-Fabrega}  \&
  {Valenzuela}}{{Colin} et~al.}{2016}]{2016arXiv160707917C}
{Colin} P.,  {Avila-Reese} V.,  {Roca-Fabrega} S.,   {Valenzuela} O.,  2016,
  preprint, \href {http://adsabs.harvard.edu/abs/2016arXiv160707917C} {}
  (\mn@eprint {arXiv} {1607.07917})

\bibitem[\protect\citeauthoryear{{Cousin}, {Buat}, {Boissier}, {Bethermin},
  {Roehlly}  \& {G{\'e}nois}}{{Cousin} et~al.}{2016}]{2016A&A...589A.109C}
{Cousin} M.,  {Buat} V.,  {Boissier} S.,  {Bethermin} M.,  {Roehlly} Y.,
  {G{\'e}nois} M.,  2016, \mn@doi [AAP] {10.1051/0004-6361/201527734}, \href
  {http://adsabs.harvard.edu/abs/2016A%26A...589A.109C} {589, A109}

\bibitem[\protect\citeauthoryear{{Creasey}, {Scannapieco}, {Nuza}, {Yepes},
  {Gottl{\"o}ber}  \& {Steinmetz}}{{Creasey}
  et~al.}{2015}]{2015ApJ...800L...4C}
{Creasey} P.,  {Scannapieco} C.,  {Nuza} S.~E.,  {Yepes} G.,  {Gottl{\"o}ber}
  S.,   {Steinmetz} M.,  2015, \mn@doi [ApJL] {10.1088/2041-8205/800/1/L4},
  \href {http://adsabs.harvard.edu/abs/2015ApJ...800L...4C} {800, L4}

\bibitem[\protect\citeauthoryear{{Dayal}, {Ferrara}  \& {Dunlop}}{{Dayal}
  et~al.}{2013}]{2013MNRAS.430.2891D}
{Dayal} P.,  {Ferrara} A.,   {Dunlop} J.~S.,  2013, MNRAS, 430, 2891

\bibitem[\protect\citeauthoryear{{De Lucia}, {Tornatore}, {Frenk}, {Helmi},
  {Navarro}  \& {White}}{{De Lucia} et~al.}{2014}]{2014MNRAS.445..970D}
{De Lucia} G.,  {Tornatore} L.,  {Frenk} C.~S.,  {Helmi} A.,  {Navarro} J.~F.,
   {White} S.~D.~M.,  2014, \mn@doi [MNRAS] {10.1093/mnras/stu1752}, \href
  {http://adsabs.harvard.edu/abs/2014MNRAS.445..970D} {445, 970}

\bibitem[\protect\citeauthoryear{{Diehl} et~al.,}{{Diehl}
  et~al.}{2006}]{2006Natur.439...45D}
{Diehl} R.,  et~al., 2006, \mn@doi [NATURE] {10.1038/nature04364}, \href
  {http://adsabs.harvard.edu/abs/2006Natur.439...45D} {439, 45}

\bibitem[\protect\citeauthoryear{{Diemand}, {Kuhlen}, {Madau}, {Zemp}, {Moore},
  {Potter}  \& {Stadel}}{{Diemand} et~al.}{2008}]{2008Natur.454..735D}
{Diemand} J.,  {Kuhlen} M.,  {Madau} P.,  {Zemp} M.,  {Moore} B.,  {Potter} D.,
    {Stadel} J.,  2008, \mn@doi [Nature] {10.1038/nature07153}, \href
  {http://adsabs.harvard.edu/abs/2008Natur.454..735D} {454, 735}

\bibitem[\protect\citeauthoryear{{Fattahi} et~al.,}{{Fattahi}
  et~al.}{2016}]{2016MNRAS.457..844F}
{Fattahi} A.,  et~al., 2016, \mn@doi [MNRAS] {10.1093/mnras/stv2970}, \href
  {http://adsabs.harvard.edu/abs/2016MNRAS.457..844F} {457, 844}

\bibitem[\protect\citeauthoryear{{Flynn}, {Holmberg}, {Portinari}, {Fuchs}  \&
  {Jahrei{\ss}}}{{Flynn} et~al.}{2006}]{2006MNRAS.372.1149F}
{Flynn} C.,  {Holmberg} J.,  {Portinari} L.,  {Fuchs} B.,   {Jahrei{\ss}} H.,
  2006, \mn@doi [MNRAS] {10.1111/j.1365-2966.2006.10911.x}, \href
  {http://adsabs.harvard.edu/abs/2006MNRAS.372.1149F} {372, 1149}

\bibitem[\protect\citeauthoryear{{Frebel} \& {Bromm}}{{Frebel} \&
  {Bromm}}{2012}]{2012ApJ...759..115F}
{Frebel} A.,  {Bromm} V.,  2012, \mn@doi [ApJ] {10.1088/0004-637X/759/2/115},
  \href {http://adsabs.harvard.edu/abs/2012ApJ...759..115F} {759, 115}

\bibitem[\protect\citeauthoryear{{Garrison-Kimmel}, {Boylan-Kolchin}, {Bullock}
   \& {Lee}}{{Garrison-Kimmel} et~al.}{2014}]{2014MNRAS.438.2578G}
{Garrison-Kimmel} S.,  {Boylan-Kolchin} M.,  {Bullock} J.~S.,   {Lee} K.,
  2014, \mn@doi [MNRAS] {10.1093/mnras/stt2377}, \href
  {http://adsabs.harvard.edu/abs/2014MNRAS.438.2578G} {438, 2578}

\bibitem[\protect\citeauthoryear{{Geen}, {Slyz}  \& {Devriendt}}{{Geen}
  et~al.}{2013}]{2013MNRAS.429..633G}
{Geen} S.,  {Slyz} A.,   {Devriendt} J.,  2013, \mn@doi [MNRAS]
  {10.1093/mnras/sts364}, \href
  {http://adsabs.harvard.edu/abs/2013MNRAS.429..633G} {429, 633}

\bibitem[\protect\citeauthoryear{{Graziani}, {Maselli}  \& {Ciardi}}{{Graziani}
  et~al.}{2013}]{2013MNRAS.431..722G}
{Graziani} L.,  {Maselli} A.,   {Ciardi} B.,  2013, \mn@doi [MNRAS]
  {10.1093/mnras/stt206}, \href
  {http://adsabs.harvard.edu/abs/2013MNRAS.431..722G} {431, 722}

\bibitem[\protect\citeauthoryear{{Graziani}, {Salvadori}, {Schneider},
  {Kawata}, {de Bennassuti}  \& {Maselli}}{{Graziani}
  et~al.}{2015}]{2015MNRAS.449.3137G}
{Graziani} L.,  {Salvadori} S.,  {Schneider} R.,  {Kawata} D.,  {de Bennassuti}
  M.,   {Maselli} A.,  2015, \mn@doi [MNRAS] {10.1093/mnras/stv494}, \href
  {http://adsabs.harvard.edu/abs/2015MNRAS.449.3137G} {449, 3137}

\bibitem[\protect\citeauthoryear{{Griffen}, {Dooley}, {Ji}, {O'Shea},
  {G{\'o}mez}  \& {Frebel}}{{Griffen} et~al.}{2016a}]{2016arXiv161100759G}
{Griffen} B.~F.,  {Dooley} G.~A.,  {Ji} A.~P.,  {O'Shea} B.~W.,  {G{\'o}mez}
  F.~A.,   {Frebel} A.,  2016a, preprint, \href
  {http://adsabs.harvard.edu/abs/2016arXiv161100759G} {} (\mn@eprint {arXiv}
  {1611.00759})

\bibitem[\protect\citeauthoryear{{Griffen}, {Ji}, {Dooley}, {G{\'o}mez},
  {Vogelsberger}, {O'Shea}  \& {Frebel}}{{Griffen}
  et~al.}{2016b}]{2016ApJ...818...10G}
{Griffen} B.~F.,  {Ji} A.~P.,  {Dooley} G.~A.,  {G{\'o}mez} F.~A.,
  {Vogelsberger} M.,  {O'Shea} B.~W.,   {Frebel} A.,  2016b, \mn@doi [ApJ]
  {10.3847/0004-637X/818/1/10}, \href
  {http://adsabs.harvard.edu/abs/2016ApJ...818...10G} {818, 10}

\bibitem[\protect\citeauthoryear{{Grogin} et~al.,}{{Grogin}
  et~al.}{2011}]{2011ApJS..197...35G}
{Grogin} N.~A.,  et~al., 2011, \mn@doi [ApJS] {10.1088/0067-0049/197/2/35},
  \href {http://adsabs.harvard.edu/abs/2011ApJS..197...35G} {197, 35}

\bibitem[\protect\citeauthoryear{{Guo}, {White}, {Li}  \&
  {Boylan-Kolchin}}{{Guo} et~al.}{2010}]{2010MNRAS.404.1111G}
{Guo} Q.,  {White} S.,  {Li} C.,   {Boylan-Kolchin} M.,  2010, \mn@doi [MNRAS]
  {10.1111/j.1365-2966.2010.16341.x}, \href
  {http://adsabs.harvard.edu/abs/2010MNRAS.404.1111G} {404, 1111}

\bibitem[\protect\citeauthoryear{{Guo} et~al.,}{{Guo} et~al.}{2016}]{Guo2016}
{Guo} Y.,  et~al., 2016, ApJ, 822, 103

\bibitem[\protect\citeauthoryear{{Hahn} \& {Abel}}{{Hahn} \&
  {Abel}}{2011}]{2011MNRAS.415.2101H}
{Hahn} O.,  {Abel} T.,  2011, \mn@doi [MNRAS]
  {10.1111/j.1365-2966.2011.18820.x}, \href
  {http://adsabs.harvard.edu/abs/2011MNRAS.415.2101H} {415, 2101}

\bibitem[\protect\citeauthoryear{{Hartwig}, {Bromm}, {Klessen}  \&
  {Glover}}{{Hartwig} et~al.}{2015}]{2015MNRAS.447.3892H}
{Hartwig} T.,  {Bromm} V.,  {Klessen} R.~S.,   {Glover} S.~C.~O.,  2015,
  \mn@doi [MNRAS] {10.1093/mnras/stu2740}, \href
  {http://adsabs.harvard.edu/abs/2015MNRAS.447.3892H} {447, 3892}

\bibitem[\protect\citeauthoryear{{Henriques}, {White}, {Thomas}, {Angulo},
  {Guo}, {Lemson}, {Springel}  \& {Overzier}}{{Henriques}
  et~al.}{2015}]{2015MNRAS.451.2663H}
{Henriques} B.~M.~B.,  {White} S.~D.~M.,  {Thomas} P.~A.,  {Angulo} R.,  {Guo}
  Q.,  {Lemson} G.,  {Springel} V.,   {Overzier} R.,  2015, \mn@doi [MNRAS]
  {10.1093/mnras/stv705}, \href
  {http://adsabs.harvard.edu/abs/2015MNRAS.451.2663H} {451, 2663}

\bibitem[\protect\citeauthoryear{{Hirschmann}, {Naab}, {Somerville}, {Burkert}
  \& {Oser}}{{Hirschmann} et~al.}{2012}]{2012MNRAS.419.3200H}
{Hirschmann} M.,  {Naab} T.,  {Somerville} R.~S.,  {Burkert} A.,   {Oser} L.,
  2012, \mn@doi [MNRAS] {10.1111/j.1365-2966.2011.19961.x}, \href
  {http://adsabs.harvard.edu/abs/2012MNRAS.419.3200H} {419, 3200}

\bibitem[\protect\citeauthoryear{{Hirschmann} et~al.,}{{Hirschmann}
  et~al.}{2013}]{2013MNRAS.436.2929H}
{Hirschmann} M.,  et~al., 2013, \mn@doi [MNRAS] {10.1093/mnras/stt1770}, \href
  {http://adsabs.harvard.edu/abs/2013MNRAS.436.2929H} {436, 2929}

\bibitem[\protect\citeauthoryear{{Hopkins}, {Kere{\v s}}, {O{\~n}orbe},
  {Faucher-Gigu{\`e}re}, {Quataert}, {Murray}  \& {Bullock}}{{Hopkins}
  et~al.}{2014}]{2014MNRAS.445..581H}
{Hopkins} P.~F.,  {Kere{\v s}} D.,  {O{\~n}orbe} J.,  {Faucher-Gigu{\`e}re}
  C.-A.,  {Quataert} E.,  {Murray} N.,   {Bullock} J.~S.,  2014, MNRAS, 445,
  581

\bibitem[\protect\citeauthoryear{{Huchra}, {Davis}, {Latham}  \&
  {Tonry}}{{Huchra} et~al.}{1983}]{1983ApJS...52...89H}
{Huchra} J.,  {Davis} M.,  {Latham} D.,   {Tonry} J.,  1983, \mn@doi [ApJS]
  {10.1086/190860}, \href {http://adsabs.harvard.edu/abs/1983ApJS...52...89H}
  {52, 89}

\bibitem[\protect\citeauthoryear{{Hunt} et~al.,}{{Hunt}
  et~al.}{2012}]{2012MNRAS.427..906H}
{Hunt} L.,  et~al., 2012, MNRAS, 427, 906

\bibitem[\protect\citeauthoryear{{Hunt}, {Dayal}, {Magrini}  \&
  {Ferrara}}{{Hunt} et~al.}{2016a}]{2016MNRAS.463.2002H}
{Hunt} L.,  {Dayal} P.,  {Magrini} L.,   {Ferrara} A.,  2016a, MNRAS, 463, 2002

\bibitem[\protect\citeauthoryear{{Hunt}, {Dayal}, {Magrini}  \&
  {Ferrara}}{{Hunt} et~al.}{2016b}]{2016MNRAS.463.2020H}
{Hunt} L.,  {Dayal} P.,  {Magrini} L.,   {Ferrara} A.,  2016b, MNRAS, 463, 2020

\bibitem[\protect\citeauthoryear{{Ibata}, {Martin}, {Irwin}, {Chapman},
  {Ferguson}, {Lewis}  \& {McConnachie}}{{Ibata}
  et~al.}{2007}]{2007ApJ...671.1591I}
{Ibata} R.,  {Martin} N.~F.,  {Irwin} M.,  {Chapman} S.,  {Ferguson} A.~M.~N.,
  {Lewis} G.~F.,   {McConnachie} A.~W.,  2007, \mn@doi [ApJ] {10.1086/522574},
  \href {http://adsabs.harvard.edu/abs/2007ApJ...671.1591I} {671, 1591}

\bibitem[\protect\citeauthoryear{{Kawata} \& {Gibson}}{{Kawata} \&
  {Gibson}}{2003}]{2003MNRAS.340..908K}
{Kawata} D.,  {Gibson} B.~K.,  2003, \mn@doi [MNRAS]
  {10.1046/j.1365-8711.2003.06356.x}, \href
  {http://adsabs.harvard.edu/abs/2003MNRAS.340..908K} {340, 908}

\bibitem[\protect\citeauthoryear{{Kawata}, {Okamoto}, {Gibson}, {Barnes}  \&
  {Cen}}{{Kawata} et~al.}{2013}]{2013MNRAS.428.1968K}
{Kawata} D.,  {Okamoto} T.,  {Gibson} B.~K.,  {Barnes} D.~J.,   {Cen} R.,
  2013, \mn@doi [MNRAS] {10.1093/mnras/sts161}, \href
  {http://adsabs.harvard.edu/abs/2013MNRAS.428.1968K} {428, 1968}

\bibitem[\protect\citeauthoryear{{Kennicutt}}{{Kennicutt}}{1998}]{1998ARA&A..36..189K}
{Kennicutt} Jr. R.~C.,  1998, \mn@doi [ARAA] {10.1146/annurev.astro.36.1.189},
  \href {http://adsabs.harvard.edu/abs/1998ARA%26A..36..189K} {36, 189}

\bibitem[\protect\citeauthoryear{{Kennicutt} \& {Evans}}{{Kennicutt} \&
  {Evans}}{2012}]{2012ARA&A..50..531K}
{Kennicutt} R.~C.,  {Evans} N.~J.,  2012, \mn@doi [ARAA]
  {10.1146/annurev-astro-081811-125610}, \href
  {http://adsabs.harvard.edu/abs/2012ARA%26A..50..531K} {50, 531}

\bibitem[\protect\citeauthoryear{{Kewley} \& {Ellison}}{{Kewley} \&
  {Ellison}}{2008}]{2008ApJ...681.1183K}
{Kewley} L.~J.,  {Ellison} S.~L.,  2008, \mn@doi [ApJ] {10.1086/587500}, \href
  {http://cdsads.u-strasbg.fr/abs/2008ApJ...681.1183K} {681, 1183}

\bibitem[\protect\citeauthoryear{{Kim} et~al.,}{{Kim}
  et~al.}{2014}]{2014ApJS..210...14K}
{Kim} J.-h.,  et~al., 2014, \mn@doi [ApJS] {10.1088/0067-0049/210/1/14}, \href
  {http://adsabs.harvard.edu/abs/2014ApJS..210...14K} {210, 14}

\bibitem[\protect\citeauthoryear{{Kirby}, {Simon}, {Geha}, {Guhathakurta}  \&
  {Frebel}}{{Kirby} et~al.}{2008}]{2008ApJ...685L..43K}
{Kirby} E.~N.,  {Simon} J.~D.,  {Geha} M.,  {Guhathakurta} P.,   {Frebel} A.,
  2008, ApJL, 685, L43

\bibitem[\protect\citeauthoryear{{Klypin}, {Trujillo-Gomez}  \&
  {Primack}}{{Klypin} et~al.}{2011}]{2011ApJ...740..102K}
{Klypin} A.~A.,  {Trujillo-Gomez} S.,   {Primack} J.,  2011, \mn@doi [ApJ]
  {10.1088/0004-637X/740/2/102}, \href
  {http://adsabs.harvard.edu/abs/2011ApJ...740..102K} {740, 102}

\bibitem[\protect\citeauthoryear{{Knebe} et~al.,}{{Knebe}
  et~al.}{2015}]{2015MNRAS.451.4029K}
{Knebe} A.,  et~al., 2015, \mn@doi [MNRAS] {10.1093/mnras/stv1149}, \href
  {http://adsabs.harvard.edu/abs/2015MNRAS.451.4029K} {451, 4029}

\bibitem[\protect\citeauthoryear{{Komatsu} et~al.,}{{Komatsu}
  et~al.}{2009}]{2009ApJS..180..330K}
{Komatsu} E.,  et~al., 2009, \mn@doi [ApJS] {10.1088/0067-0049/180/2/330},
  \href {http://adsabs.harvard.edu/abs/2009ApJS..180..330K} {180, 330}

\bibitem[\protect\citeauthoryear{{Komiya}, {Suda}, {Minaguchi}, {Shigeyama},
  {Aoki}  \& {Fujimoto}}{{Komiya} et~al.}{2007}]{2007ApJ...658..367K}
{Komiya} Y.,  {Suda} T.,  {Minaguchi} H.,  {Shigeyama} T.,  {Aoki} W.,
  {Fujimoto} M.~Y.,  2007, \mn@doi [ApJ] {10.1086/510826}, \href
  {http://adsabs.harvard.edu/abs/2007ApJ...658..367K} {658, 367}

\bibitem[\protect\citeauthoryear{{Kubryk}, {Prantzos}  \&
  {Athanassoula}}{{Kubryk} et~al.}{2015}]{2015A&A...580A.126K}
{Kubryk} M.,  {Prantzos} N.,   {Athanassoula} E.,  2015, \mn@doi [AAP]
  {10.1051/0004-6361/201424171}, \href
  {http://adsabs.harvard.edu/abs/2015A%26A...580A.126K} {580, A126}

\bibitem[\protect\citeauthoryear{{Larson}}{{Larson}}{1998}]{1998MNRAS.301..569L}
{Larson} R.~B.,  1998, \mn@doi [MNRAS] {10.1046/j.1365-8711.1998.02045.x},
  \href {http://adsabs.harvard.edu/abs/1998MNRAS.301..569L} {301, 569}

\bibitem[\protect\citeauthoryear{{Larson} et~al.,}{{Larson}
  et~al.}{2011}]{2011ApJS..192...16L}
{Larson} D.,  et~al., 2011, \mn@doi [ApJS] {10.1088/0067-0049/192/2/16}, \href
  {http://adsabs.harvard.edu/abs/2011ApJS..192...16L} {192, 16}

\bibitem[\protect\citeauthoryear{{Licquia} \& {Newman}}{{Licquia} \&
  {Newman}}{2015}]{2015ApJ...806...96L}
{Licquia} T.~C.,  {Newman} J.~A.,  2015, \mn@doi [ApJ]
  {10.1088/0004-637X/806/1/96}, \href
  {http://adsabs.harvard.edu/abs/2015ApJ...806...96L} {806, 96}

\bibitem[\protect\citeauthoryear{{Madden} et~al.,}{{Madden}
  et~al.}{2013}]{2013PASP..125..600M}
{Madden} S.~C.,  et~al., 2013, \mn@doi [PASP] {10.1086/671138}, \href
  {http://adsabs.harvard.edu/abs/2013PASP..125..600M} {125, 600}

\bibitem[\protect\citeauthoryear{{Madden} et~al.,}{{Madden}
  et~al.}{2014}]{2014PASP..126.1079M}
{Madden} S.~C.,  et~al., 2014, \mn@doi [PASP] {10.1086/679312}, \href
  {http://adsabs.harvard.edu/abs/2014PASP..126.1079M} {126, 1079}

\bibitem[\protect\citeauthoryear{{Maiolino} et~al.,}{{Maiolino}
  et~al.}{2008}]{2008A&A...488..463M}
{Maiolino} R.,  et~al., 2008, AAP, 488, 488

\bibitem[\protect\citeauthoryear{{Mannucci} et~al.,}{{Mannucci}
  et~al.}{2009}]{2009MNRAS.398.1915M}
{Mannucci} F.,  et~al., 2009, MNRAS, 398, 1915

\bibitem[\protect\citeauthoryear{{Mannucci}, {Cresci}, {Maiolino}, {Marconi}
  \& {Gnerucci}}{{Mannucci} et~al.}{2010}]{2010MNRAS.408.2115M}
{Mannucci} F.,  {Cresci} G.,  {Maiolino} R.,  {Marconi} A.,   {Gnerucci} A.,
  2010, MNRAS, 408, 2115

\bibitem[\protect\citeauthoryear{{Maselli}, {Ferrara}  \& {Ciardi}}{{Maselli}
  et~al.}{2003}]{2003MNRAS.345..379M}
{Maselli} A.,  {Ferrara} A.,   {Ciardi} B.,  2003, \mn@doi [MNRAS]
  {10.1046/j.1365-8711.2003.06979.x}, \href
  {http://adsabs.harvard.edu/abs/2003MNRAS.345..379M} {345, 379}

\bibitem[\protect\citeauthoryear{{Maselli}, {Ciardi}  \& {Kanekar}}{{Maselli}
  et~al.}{2009}]{2009MNRAS.393..171M}
{Maselli} A.,  {Ciardi} B.,   {Kanekar} A.,  2009, \mn@doi [MNRAS]
  {10.1111/j.1365-2966.2008.14197.x}, \href
  {http://adsabs.harvard.edu/abs/2009MNRAS.393..171M} {393, 171}

\bibitem[\protect\citeauthoryear{{McConnachie}}{{McConnachie}}{2012}]{2012AJ....144....4M}
{McConnachie} A.~W.,  2012, AJ, 144, 4

\bibitem[\protect\citeauthoryear{{McKee} \& {Ostriker}}{{McKee} \&
  {Ostriker}}{2007}]{2007ARA&A..45..565M}
{McKee} C.~F.,  {Ostriker} E.~C.,  2007, \mn@doi [ARAA]
  {10.1146/annurev.astro.45.051806.110602}, \href
  {http://adsabs.harvard.edu/abs/2007ARA%26A..45..565M} {45, 565}

\bibitem[\protect\citeauthoryear{{McKee} \& {Williams}}{{McKee} \&
  {Williams}}{1997}]{1997ApJ...476..144M}
{McKee} C.~F.,  {Williams} J.~P.,  1997, ApJ, \href
  {http://adsabs.harvard.edu/abs/1997ApJ...476..144M} {476, 144}

\bibitem[\protect\citeauthoryear{{McMillan}}{{McMillan}}{2011}]{2011MNRAS.414.2446M}
{McMillan} P.~J.,  2011, \mn@doi [MNRAS] {10.1111/j.1365-2966.2011.18564.x},
  \href {http://adsabs.harvard.edu/abs/2011MNRAS.414.2446M} {414, 2446}

\bibitem[\protect\citeauthoryear{{M{\'e}nard}, {Scranton}, {Fukugita}  \&
  {Richards}}{{M{\'e}nard} et~al.}{2010}]{2010MNRAS.405.1025M}
{M{\'e}nard} B.,  {Scranton} R.,  {Fukugita} M.,   {Richards} G.,  2010, MNRAS,
  405, 1025

\bibitem[\protect\citeauthoryear{{Mo}, {van den Bosch}  \& {White}}{{Mo}
  et~al.}{2010}]{2010gfe..book.....M}
{Mo} H.,  {van den Bosch} F.~C.,   {White} S.,  2010, {Galaxy Formation and
  Evolution}

\bibitem[\protect\citeauthoryear{{Monelli} et~al.,}{{Monelli}
  et~al.}{2016}]{2016ApJ...819..147M}
{Monelli} M.,  et~al., 2016, ApJ, 819, 147

\bibitem[\protect\citeauthoryear{{Mu{\~n}oz}, {Madau}, {Loeb}  \&
  {Diemand}}{{Mu{\~n}oz} et~al.}{2009}]{2009MNRAS.400.1593M}
{Mu{\~n}oz} J.~A.,  {Madau} P.,  {Loeb} A.,   {Diemand} J.,  2009, \mn@doi
  [MNRAS] {10.1111/j.1365-2966.2009.15562.x}, \href
  {http://adsabs.harvard.edu/abs/2009MNRAS.400.1593M} {400, 1593}

\bibitem[\protect\citeauthoryear{{Navarro}, {Frenk}  \& {White}}{{Navarro}
  et~al.}{1996}]{1996ApJ...462..563N}
{Navarro} J.~F.,  {Frenk} C.~S.,   {White} S.~D.~M.,  1996, \mn@doi [ApJ]
  {10.1086/177173}, \href {http://adsabs.harvard.edu/abs/1996ApJ...462..563N}
  {462, 563}

\bibitem[\protect\citeauthoryear{{Navarro} et~al.,}{{Navarro}
  et~al.}{2010}]{2010MNRAS.402...21N}
{Navarro} J.~F.,  et~al., 2010, \mn@doi [MNRAS]
  {10.1111/j.1365-2966.2009.15878.x}, \href
  {http://adsabs.harvard.edu/abs/2010MNRAS.402...21N} {402, 21}

\bibitem[\protect\citeauthoryear{{Nuza}, {Parisi}, {Scannapieco}, {Richter},
  {Gottl{\"o}ber}  \& {Steinmetz}}{{Nuza} et~al.}{2014}]{2014MNRAS.441.2593N}
{Nuza} S.~E.,  {Parisi} F.,  {Scannapieco} C.,  {Richter} P.,  {Gottl{\"o}ber}
  S.,   {Steinmetz} M.,  2014, \mn@doi [MNRAS] {10.1093/mnras/stu643}, \href
  {http://adsabs.harvard.edu/abs/2014MNRAS.441.2593N} {441, 2593}

\bibitem[\protect\citeauthoryear{{Ocvirk} et~al.,}{{Ocvirk}
  et~al.}{2014}]{2014ApJ...794...20O}
{Ocvirk} P.,  et~al., 2014, ApJ, 794, 20

\bibitem[\protect\citeauthoryear{{Papovich} et~al.,}{{Papovich}
  et~al.}{2015}]{2015ApJ...803...26P}
{Papovich} C.,  et~al., 2015, ApJ, 803, 26

\bibitem[\protect\citeauthoryear{{Patel} et~al.,}{{Patel}
  et~al.}{2013}]{2013ApJ...778..115P}
{Patel} S.~G.,  et~al., 2013, ApJ, 778, 115

\bibitem[\protect\citeauthoryear{{Peeples}, {Werk}, {Tumlinson}, {Oppenheimer},
  {Prochaska}, {Katz}  \& {Weinberg}}{{Peeples}
  et~al.}{2014}]{2014ApJ...786...54P}
{Peeples} M.~S.,  {Werk} J.~K.,  {Tumlinson} J.,  {Oppenheimer} B.~D.,
  {Prochaska} J.~X.,  {Katz} N.,   {Weinberg} D.~H.,  2014, \mn@doi [ApJ]
  {10.1088/0004-637X/786/1/54}, \href
  {http://adsabs.harvard.edu/abs/2014ApJ...786...54P} {786, 54}

\bibitem[\protect\citeauthoryear{{Pezzulli}, {Valiante}  \&
  {Schneider}}{{Pezzulli} et~al.}{2016}]{2016MNRAS.458.3047P}
{Pezzulli} E.,  {Valiante} R.,   {Schneider} R.,  2016, \mn@doi [MNRAS]
  {10.1093/mnras/stw505}, \href
  {http://adsabs.harvard.edu/abs/2016MNRAS.458.3047P} {458, 3047}

\bibitem[\protect\citeauthoryear{{Planck Collaboration} et~al.,}{{Planck
  Collaboration} et~al.}{2014}]{2014A&A...571A..16P}
{Planck Collaboration} et~al., 2014, \mn@doi [AAP]
  {10.1051/0004-6361/201321591}, \href
  {http://adsabs.harvard.edu/abs/2014A%26A...571A..16P} {571, A16}

\bibitem[\protect\citeauthoryear{{Powell}, {Slyz}  \& {Devriendt}}{{Powell}
  et~al.}{2011}]{2011MNRAS.414.3671P}
{Powell} L.~C.,  {Slyz} A.,   {Devriendt} J.,  2011, \mn@doi [MNRAS]
  {10.1111/j.1365-2966.2011.18668.x}, \href
  {http://adsabs.harvard.edu/abs/2011MNRAS.414.3671P} {414, 3671}

\bibitem[\protect\citeauthoryear{{Press} \& {Schechter}}{{Press} \&
  {Schechter}}{1974}]{1974ApJ...187..425P}
{Press} W.~H.,  {Schechter} P.,  1974, \mn@doi [ApJ] {10.1086/152650}, \href
  {http://adsabs.harvard.edu/abs/1974ApJ...187..425P} {187, 425}

\bibitem[\protect\citeauthoryear{{Robitaille} \& {Whitney}}{{Robitaille} \&
  {Whitney}}{2010}]{2010ApJ...710L..11R}
{Robitaille} T.~P.,  {Whitney} B.~A.,  2010, \mn@doi [ApJL]
  {10.1088/2041-8205/710/1/L11}, \href
  {http://adsabs.harvard.edu/abs/2010ApJ...710L..11R} {710, L11}

\bibitem[\protect\citeauthoryear{{Salvadori} \& {Ferrara}}{{Salvadori} \&
  {Ferrara}}{2009}]{salvadori09}
{Salvadori} S.,  {Ferrara} A.,  2009, MNRAS, 395, L6

\bibitem[\protect\citeauthoryear{{Salvadori} \& {Ferrara}}{{Salvadori} \&
  {Ferrara}}{2012}]{2012MNRAS.421L..29S}
{Salvadori} S.,  {Ferrara} A.,  2012, \mn@doi [MNRAS]
  {10.1111/j.1745-3933.2011.01200.x}, \href
  {http://adsabs.harvard.edu/abs/2012MNRAS.421L..29S} {421, L29}

\bibitem[\protect\citeauthoryear{{Salvadori}, {Schneider}  \&
  {Ferrara}}{{Salvadori} et~al.}{2007}]{2007MNRAS.381..647S}
{Salvadori} S.,  {Schneider} R.,   {Ferrara} A.,  2007, \mn@doi [MNRAS]
  {10.1111/j.1365-2966.2007.12133.x}, \href
  {http://adsabs.harvard.edu/abs/2007MNRAS.381..647S} {381, 647}

\bibitem[\protect\citeauthoryear{{Salvadori}, {Ferrara}, {Schneider},
  {Scannapieco}  \& {Kawata}}{{Salvadori} et~al.}{2010a}]{2010MNRAS.401L...5S}
{Salvadori} S.,  {Ferrara} A.,  {Schneider} R.,  {Scannapieco} E.,   {Kawata}
  D.,  2010a, \mn@doi [MNRAS] {10.1111/j.1745-3933.2009.00772.x}, \href
  {http://adsabs.harvard.edu/abs/2010MNRAS.401L...5S} {401, L5}

\bibitem[\protect\citeauthoryear{{Salvadori}, {Dayal}  \&
  {Ferrara}}{{Salvadori} et~al.}{2010b}]{2010MNRAS.407L...1S}
{Salvadori} S.,  {Dayal} P.,   {Ferrara} A.,  2010b, \mn@doi [MNRAS]
  {10.1111/j.1745-3933.2010.00880.x}, \href
  {http://adsabs.harvard.edu/abs/2010MNRAS.407L...1S} {407, L1}

\bibitem[\protect\citeauthoryear{{Salvadori}, {Tolstoy}, {Ferrara}  \&
  {Zaroubi}}{{Salvadori} et~al.}{2014}]{2014MNRAS.437L..26S}
{Salvadori} S.,  {Tolstoy} E.,  {Ferrara} A.,   {Zaroubi} S.,  2014, \mn@doi
  [MNRAS] {10.1093/mnrasl/slt132}, \href
  {http://adsabs.harvard.edu/abs/2014MNRAS.437L..26S} {437, L26}

\bibitem[\protect\citeauthoryear{{Salvadori}, {Sk{\'u}lad{\'o}ttir}  \&
  {Tolstoy}}{{Salvadori} et~al.}{2015}]{2015MNRAS.454.1320S}
{Salvadori} S.,  {Sk{\'u}lad{\'o}ttir} {\'A}.,   {Tolstoy} E.,  2015, \mn@doi
  [MNRAS] {10.1093/mnras/stv1969}, \href
  {http://adsabs.harvard.edu/abs/2015MNRAS.454.1320S} {454, 1320}

\bibitem[\protect\citeauthoryear{{Sarmento}, {Scannapieco}  \&
  {Pan}}{{Sarmento} et~al.}{2016}]{2016arXiv161100025S}
{Sarmento} R.,  {Scannapieco} E.,   {Pan} L.,  2016, preprint, \href
  {http://adsabs.harvard.edu/abs/2016arXiv161100025S} {} (\mn@eprint {arXiv}
  {1611.00025})

\bibitem[\protect\citeauthoryear{{Sawala} et~al.,}{{Sawala}
  et~al.}{2016}]{2016MNRAS.457.1931S}
{Sawala} T.,  et~al., 2016, \mn@doi [MNRAS] {10.1093/mnras/stw145}, \href
  {http://adsabs.harvard.edu/abs/2016MNRAS.457.1931S} {457, 1931}

\bibitem[\protect\citeauthoryear{{Scannapieco}, {Schneider}  \&
  {Ferrara}}{{Scannapieco} et~al.}{2003}]{2003ApJ...589...35S}
{Scannapieco} E.,  {Schneider} R.,   {Ferrara} A.,  2003, \mn@doi [ApJ]
  {10.1086/374412}, \href {http://adsabs.harvard.edu/abs/2003ApJ...589...35S}
  {589, 35}

\bibitem[\protect\citeauthoryear{{Scannapieco}, {Kawata}, {Brook}, {Schneider},
  {Ferrara}  \& {Gibson}}{{Scannapieco} et~al.}{2006}]{2006ApJ...653..285S}
{Scannapieco} E.,  {Kawata} D.,  {Brook} C.~B.,  {Schneider} R.,  {Ferrara} A.,
    {Gibson} B.~K.,  2006, \mn@doi [ApJ] {10.1086/508487}, \href
  {http://adsabs.harvard.edu/abs/2006ApJ...653..285S} {653, 285}

\bibitem[\protect\citeauthoryear{{Scannapieco}, {White}, {Springel}  \&
  {Tissera}}{{Scannapieco} et~al.}{2011}]{2011MNRAS.417..154S}
{Scannapieco} C.,  {White} S.~D.~M.,  {Springel} V.,   {Tissera} P.~B.,  2011,
  \mn@doi [MNRAS] {10.1111/j.1365-2966.2011.19027.x}, \href
  {http://adsabs.harvard.edu/abs/2011MNRAS.417..154S} {417, 154}

\bibitem[\protect\citeauthoryear{{Scannapieco} et~al.,}{{Scannapieco}
  et~al.}{2012}]{2012MNRAS.423.1726S}
{Scannapieco} C.,  et~al., 2012, \mn@doi [MNRAS]
  {10.1111/j.1365-2966.2012.20993.x}, \href
  {http://adsabs.harvard.edu/abs/2012MNRAS.423.1726S} {423, 1726}

\bibitem[\protect\citeauthoryear{{Scannapieco}, {Creasey}, {Nuza}, {Yepes},
  {Gottl{\"o}ber}  \& {Steinmetz}}{{Scannapieco}
  et~al.}{2015}]{2015A&A...577A...3S}
{Scannapieco} C.,  {Creasey} P.,  {Nuza} S.~E.,  {Yepes} G.,  {Gottl{\"o}ber}
  S.,   {Steinmetz} M.,  2015, \mn@doi [AAP] {10.1051/0004-6361/201425494},
  \href {http://adsabs.harvard.edu/abs/2015A%26A...577A...3S} {577, A3}

\bibitem[\protect\citeauthoryear{{Schaye} et~al.,}{{Schaye}
  et~al.}{2015}]{2015MNRAS.446..521S}
{Schaye} J.,  et~al., 2015, \mn@doi [MNRAS] {10.1093/mnras/stu2058}, \href
  {http://adsabs.harvard.edu/abs/2015MNRAS.446..521S} {446, 521}

\bibitem[\protect\citeauthoryear{{Schreiber} et~al.,}{{Schreiber}
  et~al.}{2015}]{2015A&A...575A..74S}
{Schreiber} C.,  et~al., 2015, AAP, 575, A74

\bibitem[\protect\citeauthoryear{{Simon} \& {Geha}}{{Simon} \&
  {Geha}}{2007}]{2007ApJ...670..313S}
{Simon} J.~D.,  {Geha} M.,  2007, ApJ, 670, 313

\bibitem[\protect\citeauthoryear{{Skibba} et~al.,}{{Skibba}
  et~al.}{2012}]{Skibba2012}
{Skibba} R.~A.,  et~al., 2012, ApJ, 761, 42

\bibitem[\protect\citeauthoryear{{Smith}, {Biermann}  \& {Mezger}}{{Smith}
  et~al.}{1978}]{1978A&A....66...65S}
{Smith} L.~F.,  {Biermann} P.,   {Mezger} P.~G.,  1978, AAP, \href
  {http://adsabs.harvard.edu/abs/1978A%26A....66...65S} {66, 65}

\bibitem[\protect\citeauthoryear{{Springel} et~al.,}{{Springel}
  et~al.}{2005}]{2005Natur.435..629S}
{Springel} V.,  et~al., 2005, NATURE, 435, 629

\bibitem[\protect\citeauthoryear{{Springel}, {Frenk}  \& {White}}{{Springel}
  et~al.}{2006}]{2006Natur.440.1137S}
{Springel} V.,  {Frenk} C.~S.,   {White} S.~D.~M.,  2006, \mn@doi [NATURE]
  {10.1038/nature04805}, \href
  {http://adsabs.harvard.edu/abs/2006Natur.440.1137S} {440, 1137}

\bibitem[\protect\citeauthoryear{{Springel} et~al.,}{{Springel}
  et~al.}{2008}]{2008MNRAS.391.1685S}
{Springel} V.,  et~al., 2008, \mn@doi [MNRAS]
  {10.1111/j.1365-2966.2008.14066.x}, \href
  {http://adsabs.harvard.edu/abs/2008MNRAS.391.1685S} {391, 1685}

\bibitem[\protect\citeauthoryear{{Terrazas}, {Bell}, {Henriques}  \&
  {White}}{{Terrazas} et~al.}{2016}]{2016MNRAS.459.1929T}
{Terrazas} B.~A.,  {Bell} E.~F.,  {Henriques} B.~M.~B.,   {White} S.~D.~M.,
  2016, \mn@doi [MNRAS] {10.1093/mnras/stw673}, \href
  {http://adsabs.harvard.edu/abs/2016MNRAS.459.1929T} {459, 1929}

\bibitem[\protect\citeauthoryear{{Tolstoy}, {Hill}  \& {Tosi}}{{Tolstoy}
  et~al.}{2009}]{2009ARA&A..47..371T}
{Tolstoy} E.,  {Hill} V.,   {Tosi} M.,  2009, ARAA, 47, 371

\bibitem[\protect\citeauthoryear{{Tumlinson}}{{Tumlinson}}{2006}]{2006ApJ...641....1T}
{Tumlinson} J.,  2006, \mn@doi [ApJ] {10.1086/500383}, \href
  {http://adsabs.harvard.edu/abs/2006ApJ...641....1T} {641, 1}

\bibitem[\protect\citeauthoryear{{Tumlinson}}{{Tumlinson}}{2010}]{2010ApJ...708.1398T}
{Tumlinson} J.,  2010, \mn@doi [ApJ] {10.1088/0004-637X/708/2/1398}, \href
  {http://adsabs.harvard.edu/abs/2010ApJ...708.1398T} {708, 1398}

\bibitem[\protect\citeauthoryear{{Valiante}, {Schneider}, {Salvadori}  \&
  {Bianchi}}{{Valiante} et~al.}{2011}]{2011MNRAS.416.1916V}
{Valiante} R.,  {Schneider} R.,  {Salvadori} S.,   {Bianchi} S.,  2011, \mn@doi
  [MNRAS] {10.1111/j.1365-2966.2011.19168.x}, \href
  {http://adsabs.harvard.edu/abs/2011MNRAS.416.1916V} {416, 1916}

\bibitem[\protect\citeauthoryear{{Valiante}, {Schneider}, {Salvadori}  \&
  {Gallerani}}{{Valiante} et~al.}{2014}]{2014MNRAS.444.2442V}
{Valiante} R.,  {Schneider} R.,  {Salvadori} S.,   {Gallerani} S.,  2014,
  \mn@doi [MNRAS] {10.1093/mnras/stu1613}, \href
  {http://adsabs.harvard.edu/abs/2014MNRAS.444.2442V} {444, 2442}

\bibitem[\protect\citeauthoryear{{Valiante}, {Schneider}, {Volonteri}  \&
  {Omukai}}{{Valiante} et~al.}{2016a}]{2016MNRAS.457.3356V}
{Valiante} R.,  {Schneider} R.,  {Volonteri} M.,   {Omukai} K.,  2016a, \mn@doi
  [MNRAS] {10.1093/mnras/stw225}, \href
  {http://adsabs.harvard.edu/abs/2016MNRAS.457.3356V} {457, 3356}

\bibitem[\protect\citeauthoryear{{Valiante}, {Schneider}, {Volonteri}  \&
  {Omukai}}{{Valiante} et~al.}{2016b}]{valiante16}
{Valiante} R.,  {Schneider} R.,  {Volonteri} M.,   {Omukai} K.,  2016b, MNRAS,
  457, 3356

\bibitem[\protect\citeauthoryear{{Vincenzo}, {Matteucci}, {Vattakunnel}  \&
  {Lanfranchi}}{{Vincenzo} et~al.}{2014}]{2014MNRAS.441.2815V}
{Vincenzo} F.,  {Matteucci} F.,  {Vattakunnel} S.,   {Lanfranchi} G.~A.,  2014,
  \mn@doi [MNRAS] {10.1093/mnras/stu710}, \href
  {http://adsabs.harvard.edu/abs/2014MNRAS.441.2815V} {441, 2815}

\bibitem[\protect\citeauthoryear{{Vogelsberger} et~al.,}{{Vogelsberger}
  et~al.}{2014}]{2014Natur.509..177V}
{Vogelsberger} M.,  et~al., 2014, \mn@doi [NATURE] {10.1038/nature13316}, \href
  {http://adsabs.harvard.edu/abs/2014Natur.509..177V} {509, 177}

\bibitem[\protect\citeauthoryear{{Wang} et~al.,}{{Wang}
  et~al.}{2011}]{2011MNRAS.413.1373W}
{Wang} J.,  et~al., 2011, \mn@doi [MNRAS] {10.1111/j.1365-2966.2011.18220.x},
  \href {http://adsabs.harvard.edu/abs/2011MNRAS.413.1373W} {413, 1373}

\bibitem[\protect\citeauthoryear{{Wang}, {Han}, {Cooper}, {Cole}, {Frenk}  \&
  {Lowing}}{{Wang} et~al.}{2015}]{2015MNRAS.453..377W}
{Wang} W.,  {Han} J.,  {Cooper} A.~P.,  {Cole} S.,  {Frenk} C.,   {Lowing} B.,
  2015, \mn@doi [MNRAS] {10.1093/mnras/stv1647}, \href
  {http://adsabs.harvard.edu/abs/2015MNRAS.453..377W} {453, 377}

\bibitem[\protect\citeauthoryear{{Wetzel}, {Deason}  \&
  {Garrison-Kimmel}}{{Wetzel} et~al.}{2015}]{2015ApJ...807...49W}
{Wetzel} A.~R.,  {Deason} A.~J.,   {Garrison-Kimmel} S.,  2015, ApJ, 807, 49

\bibitem[\protect\citeauthoryear{{White} \& {Frenk}}{{White} \&
  {Frenk}}{1991}]{1991ApJ...379...52W}
{White} S.~D.~M.,  {Frenk} C.~S.,  1991, \mn@doi [ApJ] {10.1086/170483}, \href
  {http://adsabs.harvard.edu/abs/1991ApJ...379...52W} {379, 52}

\bibitem[\protect\citeauthoryear{{Yepes}, {Gottl{\"o}ber}  \&
  {Hoffman}}{{Yepes} et~al.}{2014}]{2014NewAR..58....1Y}
{Yepes} G.,  {Gottl{\"o}ber} S.,   {Hoffman} Y.,  2014, \mn@doi [NAR]
  {10.1016/j.newar.2013.11.001}, \href
  {http://adsabs.harvard.edu/abs/2014NewAR..58....1Y} {58, 1}

\bibitem[\protect\citeauthoryear{{de Bennassuti}, {Schneider}, {Valiante}  \&
  {Salvadori}}{{de Bennassuti} et~al.}{2014}]{2014MNRAS.445.3039D}
{de Bennassuti} M.,  {Schneider} R.,  {Valiante} R.,   {Salvadori} S.,  2014,
  \mn@doi [MNRAS] {10.1093/mnras/stu1962}, \href
  {http://adsabs.harvard.edu/abs/2014MNRAS.445.3039D} {445, 3039}

\bibitem[\protect\citeauthoryear{{de Bennassuti}, {Salvadori}, {Schneider},
  {Valiante}  \& {Omukai}}{{de Bennassuti} et~al.}{2016}]{2016arXiv161005777D}
{de Bennassuti} M.,  {Salvadori} S.,  {Schneider} R.,  {Valiante} R.,
  {Omukai} K.,  2016, preprint, \href
  {http://adsabs.harvard.edu/abs/2016arXiv161005777D} {} (\mn@eprint {arXiv}
  {1610.05777})

\bibitem[\protect\citeauthoryear{{van Dokkum} et~al.,}{{van Dokkum}
  et~al.}{2013}]{2013ApJ...771L..35V}
{van Dokkum} P.~G.,  et~al., 2013, ApJ, 771, L35

\makeatother
\end{thebibliography}

\begin{appendix}

\section{MW Halo properties and comparison with other zoom-in simulations}

\begin{table}

%%\begin{centering}

\begin{tabular}{|c|c|c|c|}
\hline
Simulation & Halo ID & Pair & Cosmology \tabularnewline
\hline 
\hline 
GAMESH & MW & no & \text{Planck} \tabularnewline
\hline
AQUARIUS & Aq-A & no & \text{WMAP-5} \tabularnewline
\hline
AQUARIUS & Aq-C & no & \text{WMAP-5} \tabularnewline
\hline
ELVIS & Hamilton & yes & \text{WMAP-7} \tabularnewline
\hline
ELVIS & iHamilton & no & \text{WMAP-7} \tabularnewline
\hline
ELVIS & Hall & yes & \text{WMAP-7} \tabularnewline
\hline
ELVIS & iHall & no & \text{WMAP-7} \tabularnewline
\hline
ELVIS & iHall HiRes& no & \text{WMAP-7} \tabularnewline
\hline
CATERPILLAR & Cat-8/LX13 & no & \text{Planck} \tabularnewline
\hline
CATERPILLAR & Cat-8/LX14 & no & \text{Planck} \tabularnewline
\hline
CATERPILLAR & Cat-12/LX13 & no & \text{Planck} \tabularnewline
\hline
CATERPILLAR & Cat-12/LX14 & no & \text{Planck} \tabularnewline
\hline
APOSTLE/DOVE & Ap-8 & yes & \text{WMAP-7} \tabularnewline
\hline
\end{tabular}

\par\
%%end{centering}
\caption{Summary of reference zoom-in simulations adopted for comparison with the MW halo found in \texttt{GAMESH}: AQUARIUS \citep{2008MNRAS.391.1685S}, ELVIS \citep{2014MNRAS.438.2578G}, CATERPILLAR \citep{2016ApJ...818...10G,2016arXiv161100759G}, APOSTLE \citep{2016MNRAS.457..844F, 2016MNRAS.457.1931S}.  In this table Halo ID is the unique name of the halo as found in the cited literature, while the column named Pair indicates if the halo is found in a binary configuration with a M31 analogue or it is an isolated one. The Planck cosmology refers to \citet{2014A&A...571A..16P} while WMAP-5 to \citet{2009ApJS..180..330K} and WMAP-7 to \citet{2011ApJS..192...16L}.}

\label{table:DMSims}
\end{table}

\begin{table*}

%%\begin{centering}

\begin{tabular}{|c|c|c|c|c|c|c|c|c|}
\hline
Halo ID & m$_p$~$[M_{\odot}]$ & M$_{\texttt{vir}}$~$[M_{\odot}]$ & R$_{\texttt{vir}}$~[kpc] & R$_s$ [Mpc] & $z_{0.5}$ & $c$ & V$_\texttt{max}$~[km/s] & R$_\texttt{max}$ [kpc] \tabularnewline
\hline 
\hline 
MW & $3.38 \times 10^5$ & $1.72 \times 10^{12}$ & 317 & 2.0 & 1.46 & 12.7 & 198.19 & 56.5 \tabularnewline
\hline
Aq-A-4 & 3.93$\times10^5$ & $1.84\times10^{12}$ & 246 & - & 1.93 & 16.21 & 209.24 & 28.2 \tabularnewline
\hline
Aq-A-2 & 1.37$\times10^4$ & $1.84\times10^{12}$ & 246 & - & 1.93 & 16.19 & 208.49 & 28.1 \tabularnewline
\hline
Aq-C-4 & 3.21$\times10^5$ & $1.79\times10^{12}$ & 244 & - & 2.23 & 14.84 & 223.20 & 33.6 \tabularnewline
\hline
Aq-C-2 & 1.40$\times10^4$ & $1.77\times10^{12}$ & 243 & - & 2.23 & 15.21 & 222.40 & 32.5 \tabularnewline
\hline
Hamilton& 1.90$\times10^5$ & $1.76\times10^{12}$ & 315 & 1.39 & 1.47 & 9.9 & 197 & - \tabularnewline
\hline
iHamilton & 1.90$\times10^5$ & $1.86\times10^{12}$ & 321 & 1.55 & 2.11 & 14.2 & 203 & - \tabularnewline
\hline
Hall & 1.90$\times10^5$ & $1.52\times10^{12}$ & 299 & 1.35 & 1.04 & 10.3 & 180 & - \tabularnewline
\hline
iHall & 1.90$\times10^5$ & $1.71\times10^{12}$ & 300 & 1.59 & 1.13 & 6.0 & 172 & - \tabularnewline
\hline
iHall HiRes & 2.35$\times10^4$ & $1.67\times10^{12}$ & 309 & 1.59 & 1.07 & 5.8 & 167 & - \tabularnewline
\hline
Cat-8/LX13 & 2.39$\times10^5$ & $1.70\times10^{12}$ & 315 & 1.55 & 1.52 & 13.3 & 197.64 & 39.81  \tabularnewline
\hline
Cat-8/LX14 & 2.98$\times10^4$ & $1.70\times10^{12}$ & 315 & 1.54 & 1.52 & 13.5 & 198.56 & 40.82  \tabularnewline
\hline
Cat-12/LX13 & 2.39$\times10^5$ & $1.77\times10^{12}$ & 319 & 1.239 & 1.37 & 11.7 & 191.32 & 49.44 \tabularnewline
\hline
Cat-12/LX14 & 2.98$\times10^4$ & $1.76\times10^{12}$ & 319 & 1.162 & 1.37 & 11.4 & 191.30 & 52.72 \tabularnewline
\hline
Ap-8 & 5.0$\times10^5$/8.8$\times10^6$ & $1.72\times10^{12}$ & - & - & - & - & - & - \tabularnewline
\hline
\end{tabular}

\par\
%%end{centering}
\caption{Summary of Milky Way-like halo properties taken for comparison with \texttt{GAMESH} (see Table~\ref{table:DMSims}). Here  Halo ID is the unique name of the halo, m$_p$ is the DM particle mass adopted for the highest resolution run, M$_{\texttt{vir}}$, R$_{\texttt{vir}}$ are virial mass and radius respectively, R$_s$ is the maximum radius of the sphere non contaminated by lower resolution particles (Note that in \texttt{GAMESH} this value refers to half of the side length of the cube contaminated only by high-resolution particles.), $z_{0.5}$ is the halo formation redshift (see text for more details). The concentration parameter $c$ is calculated following \citealt{1996ApJ...462..563N, 2010MNRAS.402...21N}, while V$_\texttt{max}$ and R$_\texttt{max}$ are computed from the rotation curve of the halo (see \citealt{2014MNRAS.438.2578G} for a definition). Finally, note that some values are not found in the reference literature of the AQUARIUS, ELVIS and APOSTLE simulations.}
\label{table:DMHalos}
\end{table*}

In this appendix we summarize the structural properties of the MW and compare with similar halos found in the AQUARIUS, ELVIS, CATERPILLAR and APOSTLE simulations. Table~\ref{table:DMSims} and Table~\ref{table:DMHalos} collect halos having M$_\texttt{vir} \sim 1.5-1.9\times 10^{12}$~M$_{\odot}$ and summarize their structural properties. The Hamilton/Hall halos of the ELVIS catalogue and Cat-8/Cat-12 in the CATERPILLAR have been selected as comparison targets from the available statistical samples, while the properties of Aq-A and Aq-C will be discussed later, when comparing 
their accretion history to the MW one. We also included halo Ap-8 from the APOSTLE project even if this is a hydrodynamical simulation and many structural properties of Ap-8 are not documented. Its satellites distribution, on the other hand, has been extensively studied in the literature and will allow future dedicated comparisons (Mancini et al. in prep).
 
From Table~\ref{table:DMHalos} we infer that our simulation adopts a mass resolution and cosmology compatible with level 13 (Halos Cat-8/LX13 and Cat-12/LX13) in the CATERPILLAR project, while the ELVIS simulation provides halos better resolved by a factor 1.8. The latter two simulations have released halo catalogues with a mass resolution increased roughly by one order of magnitude: 
iHall HighRes and Cat-8/LX14, Cat-12/LX14.
As for the adopted cosmology, our simulation and the CATERPILLAR suite adopt the same Planck parameters while ELVIS and APOSTLE rely on WMAP-7 measurements. Finally, 
ELVIS has simulated both isolated (iHamilton/iHall) and paired halos (with an M31 analogue at a distance of $\sim 800$~kpc), while the MW halo and the halos identified in the 
CATERPILLAR sample are isolated.

By checking the values in Table~\ref{table:DMHalos} it is evident that the MW halo structural properties, such as its virial radius R$_\texttt{vir}$, concentration parameter $c$ (see \citealt{1996ApJ...462..563N, 2010MNRAS.402...21N} for a definition) and maximum circular velocity V$_\texttt{max}$ (\citealt{2014MNRAS.438.2578G}) are compatible with the scatter in the Cat-8/Cat-12 at level 13 and iHamilton/iHall. Interestingly we note that the larger scatter in halo structural properties($c$, R$_\texttt{vir}$, R$_\texttt{max}$) can be linked to their peculiar accretion histories (see  section~\ref{subsec:MWassembly} for more details). 

The ability of these simulations to resolve the halo environment is quantified by R$_s$, the size of the "non contaminated region". In our simulation, a cubic volume of  4 cMpc side-length
is uncontaminated,  which is larger that spherical regions with R$_s$ = 1.1-1.5 cMpc surrounding the other halos. 
This rich (but still computationally affordable) dynamical information will allow future studies to adopt accurate RT feedback and/or particle-tagging techniques to model 
 in-homogeneous enrichment and to trace the distribution of stellar populations inside the MW halo.
 
\section{Comparison with other SAM predictions}

Here we briefly compare our predictions on the redshift evolution of SFR, M$_{\star}$ and M$_{Z}$, with independent SAM models coupled to DM simulations. 
As often discussed in many SAM comparisons projects \citep{2015MNRAS.451.4029K,2012MNRAS.419.3200H} a clear comparison should be based on a set of similar or, at least, controlled physical assumptions. This is largely beyond the scope of this paper because our model is based on single DM halo simulation and does not provide any statistical prediction across the natural diversity in the growth of DM halos \citep{2010MNRAS.402...21N,2016MNRAS.459.1929T}. Despite this intrinsic limitation, when compared with predictions based on Milky Way-like halos having similar mass, structural properties and smooth accretion history, \texttt{GAMESH} results are in broad agreement with the mean SFR and M$_{\star}$ shown in \citet{2013MNRAS.436.2929H} (see top left panels of Figure 2 and 5 relative to $12 < {\rm log} M < 13$) or with the FULL case of Figure 6 in \citet{2012MNRAS.419.3200H}. Note that these models are not tuned to reproduce the MW properties at $z=0$ (as imposed in \texttt{GAMESH}) and then their final values can be sensitively different. 

Once \texttt{GAMESH} is compared with predictions based on Aq-A-5, Aq-C-5 \citep{2011MNRAS.417..154S}, our model agrees with the  M$_{\star}$ evolution even if the intrinsic differences in the SAM assumptions and parameter tuning cannot simultaneously guarantee a similar trend in SFR. This is also the case for Aq-A-3 discussed in \citet{2014MNRAS.445..970D}. 

An excellent agreement is found instead with the model $m_3$ of \citet{2016A&A...589A.109C}, mainly because it adopts a similar description of star formation and metal enrichment. 
These authors use in fact an identical prescription to compute the SFR and adopt an efficiency $\epsilon_{\star} = 0.02$ on a set of 56 MW-like halos having $3 < M_{\star}/10^{10} M_{\odot} < 7$ and $0.7 < M_{\rm DM}/10^{12} M_{\odot} < 3$ (in their jargon $MW-sister$ and $MW-cousins$ halos). They also find  SFR$(z=0) = (1- 5)$~M$_{\odot}/yr$. Despite their disk modeling is not implemented in \texttt{GAMESH}, our MW halo can be certainly classified as a $MW-cousin$ and the comparison of  model predictions for the SFR is simply straightforward (compare Figure 6 with Figure 9 in their paper).

\end{appendix}
\label{lastpage}
\end{document}